\documentclass{jfm}

\usepackage{xcolor}
\usepackage{comment} 
\usepackage{subcaption}
\usepackage{amsmath}
\usepackage{siunitx}

\begin{document}

\newtheorem{lemma}{Lemma}
\newtheorem{corollary}{Corollary}

\shorttitle{Available energy including momentum constraints and eddy energy} 
%for header on odd pages
\shortauthor{Tailleux \& Harris} %for header on even pages

%\title{On the available energy of axisymmetric compressible stratified %vortex motions and generalised buoyancy/inertial forces}

\title{The generalised buoyancy/inertial forces and available energy of axisymmetric compressible stratified vortex motions} 

\author
 {
 R\'emi Tailleux\aff{1}
  \corresp{\email{R.G.J.Tailleux@reading.ac.uk}},
  \and 
  Bethan L. Harris\aff{1}\aunote{Current affiliation: UK Centre for Ecology \& Hydrology, Wallingford, United Kingdom.}
  }

\affiliation
{
\aff{1}
Department of Meteorology, University of Reading, Reading, RG6 6BB, United Kingdom
}

\maketitle

\begin{abstract}

\textcolor{blue}
{Adiabatic and inviscid axisymmetric perturbations to a stable reference vortex in gradient wind balance} \textcolor{violet}{are known} \textcolor{blue}{to experience two kinds of restoring forces: one that is proportional to both the perturbation density and the reference pressure gradient, and one that is purely radial and proportional to the squared angular momentum perturbation. We show that the work required to move a fluid parcel against such forces from its equilibrium to actual position is path-independent and formally equivalent to the available energy accounting for momentum constraints previously constructed by Andrews (2006) and Codoban and Shepherd (2006). Physically, this work represents the energy of the unbalanced part of the vortex and hence a form of eddy energy. It can be partitioned into available acoustic energy, slantwise available potential energy and centrifugal potential energy\textcolor{violet}{. We} show that the conditions required for these energies to be positive definite correspond to the classical conditions for symmetric stability, valid for both small and finite-amplitude perturbations.} 
\begin{comment} \textcolor{violet}{Using an intensifying warm core cyclonic vortex as an illustration, our framework shows that while a positive diabatic source of heating is the ultimate source of energy for both the balanced and balanced parts of the vortex, friction is essential for the intensification of the balanced part while at the same time an overall sink of energy. It also shows that the intensifications of the balanced and unbalanced parts have distinct energy signatures and that the generalised inertial/buoyancy forces pertain only to the intensification of the unbalanced part.} 
\end{comment} 
\textcolor{violet}{
The new available energy framework possesses various advantageous features over previous approaches that shed new light on how the thermodynamic and mechanical sinks/sources of energy control the intensification of the balanced and unbalanced parts of a warm core cyclonic vortex. These features should prove useful in the future to re-examine and clarify the links between the different existing paradigms of tropical cyclone (TC) intensification within a single unifying theoretical framework. } 
\begin{comment}
\textcolor{blue}{
The new available energy framework naturally defines thermodynamic and mechanical efficiencies, \textcolor{violet}{which quantify} the fraction of the thermodynamic and mechanical sources/sinks of energy affecting the balanced and unbalanced parts of the vortex\textcolor{violet}{. It also} clarifies how these are linked to the forces driving the secondary circulation.} 
\textcolor{violet}{Moreover, this framework contains many of the physical ingredients of existing paradigms of tropical cyclone (TC) intensification as well as new ones, which should prove useful for unifying theories of TC intensification in the future.}
\end{comment} 
\end{abstract}

\section{Introduction}

The concepts of buoyancy and buoyancy forces are central to the theoretical description of most observed phenomena in stratified fluids \citep{Turner1973}. The conventional buoyancy force is generally \textcolor{blue}{envisioned} as a purely vertical force due to the restoring effect of gravity that a fluid parcel experiences \textcolor{blue}{as} it works against the environment. Its magnitude is determined in large part by the background vertical gradient and near material invariance of specific entropy, of which the main relevant measure is the squared buoyancy frequency profile $N_r^2(z)$. It is typically introduced by viewing the total pressure and density fields as perturbations to background reference hydrostatic pressure and density fields $p_r(z)$ and $\rho_r(z)$ that are functions of height $z$ alone. Unless the fluid is close to a state of rest, \textcolor{blue}{there is some ambiguity about how best to define the reference state characterising the environment, and hence how to define buoyancy. This ambiguity extends to energetics and the concept of available potential energy (APE) density, which physically represents} the work against such a buoyancy force needed to move a fluid parcel adiabatically from its equilibrium position in the notional reference state to its actual position \citep{Andrews1981,Holliday1981,Tailleux2013b,Tailleux2018}. 
%The APE density, like the conventional buoyancy, can be defined for %arbitrarily specified hydrostatic reference pressure and density fields %$p_r(z)$ and $\rho_r(z)$ (which can also be assumed to be time dependent if %needed). 
\textcolor{blue}{For a simple fluid (i.e., one whose equation of state depends only on temperature and pressure), the volume integral of the APE density coincides with \citet{Lorenz1955}'s globally defined APE only if the reference state is defined as the state of minimum potential energy obtainable from the actual state by means of an adiabatic re-arrangement of mass. For reference states defined differently (for instance, if $\rho_r(z)$ is defined in terms of the horizontally-averaged density field),} \textcolor{violet}{ \citet{Andrews1981} and \citet{Tailleux2013b} have shown that the integral of the APE density generally exceeds the Lorenz APE.} In the case of a more complex fluid, such as a moist atmosphere or salty ocean, the issue is considerably more complicated --- see \citet{Saenz2015,Hieronymus2015,Wong2016,Stansifer2017,Harris2018} for discussions of some of the issues involved. 

\textcolor{blue}{One of the main reasons that the concept of buoyancy force plays such a central role in the study of stratified fluids is \textcolor{violet}{that} for fluid flows close to their} \textcolor{violet}{resting} \textcolor{blue}{equilibrium position, the buoyancy force is in general a very good predictor of the total force exerted on fluid parcels, thus allowing the use of the so-called parcel method} \textcolor{violet}{(see \citet{Thorpe1989} for a wide ranging review of the method and of its limitations for the discussion of various types of instabilities)}. \textcolor{blue}{The parcel method, when applicable, represents a powerful and simplifying tool for analysing the stability of fluids, as it makes it possible to describe the fluid behaviour in terms of ordinary differential equations (ODEs) instead of PDEs, due to the buoyancy force being local in nature \citep{Bannon2004}. }

\textcolor{blue}{For fluid flows far away from their equilibrium position, however, the conventional buoyancy force generally stops being a good predictor of the total force exerted on fluid parcel, thus rendering the use of the parcel method problematic.}
\begin{comment} 
The parcel method is so useful and attractive (at least, to theoreticians), however, that before giving up on using it, one may try alternative strategies to try to make it work. Thus, if one insists on using a reference state function of $z$, one may try to redefine $\rho_r(z)$ and $p_r(z)$ so the relative contribution of the associated buoyancy force to the total force exerted on fluid parcels is \textcolor{violet}{maximised. Although} the solution to this problem is unknown, there is no reason to assume that such fields would necessarily coincide with \citet{Lorenz1955}'s reference fields. However, even if the optimal choice of $\rho_r(z)$ were known, the conventional force might still perform too poorly to be of much use for using the parcel method. 
In that case, the only viable alternative is to give up on using a reference state function of $z$ only. 
\end{comment} 
\textcolor{blue}{
This naturally has led to approaches redefining buoyancy relative to more locally-defined reference pressure and density fields, which are more representative of the actual pressure and density fields. One key example is} the concept of Convective Available Potential Energy (CAPE) \citep{Moncrieff1976}, which refers to the total work done by the fluid parcel buoyancy --- defined relative to a local atmospheric sounding --- from its level of free convection (LFC) to an upper level of neutral buoyancy (LNB); in this case, \textcolor{blue}{the fluid parcel buoyancy also depends on total water content.} In the oceanic context, \citet{McDougall1987} proposed a generalisation of the concept of buoyancy valid for lateral displacements $\delta {\bf x}$, for which he derived the expression $b = - {\bf N} \cdot \delta {\bf x}$, where ${\bf N} = g(\alpha \nabla \theta - \beta \nabla S)$ is the so-called neutral vector, whose vertical component is equal to the locally defined squared buoyancy frequency. In the context of stratified turbulent mixing, \citet{Arthur2017} recently discussed some of the consequences for the estimation of turbulent mixing of using a globally- versus locally-defined reference state. In that case, the issue arises from the difficulty of connecting the study of turbulent mixing in small domains --- for which the reference state is naturally defined in terms of a global horizontal average or adiabatically sorted state --- with that in the field, which is usually based on the use of a locally-defined reference state, e.g. \citet{Thorpe1977}. 

\textcolor{blue}{To distinguish between the two approaches, \citet{Smith2005} introduced the concepts of `system' versus `local' buoyancy to refer to the buoyancy defined relative to a resting and non-resting state respectively. Recognising such a distinction appears to be crucial for resolving the controversy about the role of buoyancy in tropical cyclones, as the literature generally finds buoyancy relevant only if defined relative to a non-resting state. In contrast to the `system' buoyancy, the local buoyancy force (called generalised buoyancy force in \citet{Smith2005}) is directed along the pressure gradient of the non-resting reference state, and therefore possesses both radial and vertical \textcolor{violet}{components. For} positive buoyancy anomalies, these point \textcolor{violet}{inwards} and \textcolor{violet}{upwards} respectively. The concept of generalised buoyancy force therefore naturally clarifies how thermodynamic sources/sinks of energy drive the secondary circulation, as pointed out in \citet{Montgomery2017}'s recent review.} \textcolor{blue}{In the context of the local theory of available potential energy (APE) (see \citet{Tailleux2013} for a review of the topic), \textcolor{violet}{recognising the distinction between system and local buoyancy paved the way} for understanding how to partition the finite-amplitude local APE density into mean and eddy components, the mean and eddy APE being thus related to the work done by the former and the latter respectively \citep{Scotti2014,Novak2018,Tailleux2018}}. 

\textcolor{violet}{The generalised buoyancy force directed along the mean pressure gradient and proportional to the density anomaly --- as discussed by \citet{Smith2005} --- is only one of the restoring (or destabilising as the case may be) forces acting on a fluid parcel. Another type of force (also alluded to by \citet{Smith2005}, see their Section 2.3) is the inertial/centrifugal force first discussed by \citet{Rayleigh1916}, which arises from angular momentum conservation constraints. In this paper, we show that the centrifugal force plays a key role in the construction of the available energetics of the unbalanced part of the vortex. We also show that it imposes some important constraints on the magnitude of the diabatic heating necessary to sustain the generalised buoyancy force discussed by \citet{Smith2005}.}  

\textcolor{violet}{If one accepts that the standard APE represents the work done by the standard buoyancy force, one may legitimately ask whether the problem of how to generalise the concept of APE accounting for momentum constraints --- a long-standing issue in the field --- could be solved by simply linking it to the work done by the generalised buoyancy and inertial/centrifugal forces. It should be possible to define the work done by the generalised buoyancy force in terms of the existing expression for the local APE density, viz,}
\textcolor{blue}{
\begin{equation}
    \Pi = h(\eta,S,p)-h(\eta,S,p_r(z_r)) + g(z-z_r) + (p_r(z)-p)/\rho,
    \label{PI_expression} 
\end{equation}
}
\textcolor{violet}{since such an expression does not actually require the reference pressure $p_r$ to be a function of z only.} \textcolor{blue}{Here, $h$ is the specific enthalpy, $\eta$ the specific entropy, $S$ chemical composition, $p$ pressure, $\rho$ density, and $z_r$ the reference position of the fluid parcel in the reference state \citep{Tailleux2018}.} \textcolor{violet}{Such a property was previously noted by \citet{Tailleux2018}, who suggested that defining $\Pi$ as per Eq. (\ref{PI_expression}) in terms of a more general reference pressure field $p_m(x,y,z,t)$ (that is, one additionally depending on horizontal position and time) would naturally provide a definition of `eddy' APE density $\Pi_e$.} \textcolor{blue}{Doing so, however, introduces a term proportional to the horizontal pressure gradient $\nabla_h p_m$ in the budget for the sum of the kinetic energy and $\Pi_e$, which no longer closes.} \textcolor{blue}{In this paper, we investigate the possibility --- first suggested by \citet{Cho1993} --- of achieving a closed energy budget by incorporating the work of the centrifugal forces into the definition of available energy. This is made possible by the fact that} 
the inertial/centrifugal forces experienced by displaced fluid parcels have the nice property of being purely radial and proportional to the squared angular momentum anomaly, as previously shown and discussed by \citet{Rayleigh1916} and \citet{Emanuel1994}. \textcolor{blue}{As it turns out, the sum of the APE and centrifugal potential energy} defines an eddy form of available energy that is formally similar to that previously constructed for a zonal flow by \citet{Codoban2003} and for axisymmetric vortex motions by \citet{Codoban2006} and \citet{Andrews2006} (CSA06 thereafter). Whereas CSA06's approach emphasises the Energy-Casimir method \citep{Haynes1988,Shepherd1993}, our approach emphasises a force-based viewpoint of available energetics, which we think is much simpler and more intuitive, as well as much less abstract. 

\textcolor{blue}{An important result of \citet{Codoban2003,Codoban2006}} \textcolor{violet}{and} \textcolor{blue}{\citet{Andrews2006} is that the available energy of axisymmetric perturbations to the background reference state is conserved independently from that of the energy of the background reference state, a result also established in the present paper.} \textcolor{violet}{This result is important because it implies the impossibility of reversible energy conversions between the balanced and balanced parts of the vortex, in contrast to what is known to be the case in the full three-dimensional case. As a result, it is not possible for the `eddy' energy of the unbalanced part to reversibly alter the `mean' energy of the balanced vortex, nor is it possible for the `mean' APE of the balanced part of the vortex to be converted into the `eddy' APE and KE of the unbalanced part, the latter characterising the energy pathways of baroclinic instability that dominate Lorenz energy cycle. Whether significant differences exist between axisymmetric and asymmetric TC evolution has been a longstanding question in the TC literature, which is at the centre of the rotating convection paradigm, see \citet{Persing2013}, \citet{Montgomery2014,Montgomery2017}. Strong differences in permissible energy conversions represent an important difference between the axisymmetric and asymmetric cases that has not received much attention so far, but which might prove useful as a way to characterise the rotating convection paradigm in the future. }
  
This paper is organised as follows. Section 2 describes the model formulation. Section 3 details the construction of the available energy, and provides illustrations of the new energetic concepts for an analytical axisymmetric vortex in a dry atmosphere. Section 4 demonstrates the potential usefulness of the framework by discussing the energetics of the growth and decay of an axisymmetric vortex due to diabatic heating, in the presence of viscous effects. Section 5 discusses the results and connections with the TC intensification literature.

\section{Model formulation}
\label{model_formulation}

The evolution of compressible vortex motions is most usefully described by writing the Navier-Stokes equations in cylindrical coordinates $(r,\phi,z)$:
\begin{equation} 
 \frac{Du}{Dt} - \left ( f + \frac{v}{r} \right ) v + \nu \frac{\partial p}{\partial r} = D_u ,
 \label{radial}
\end{equation}
\begin{equation}
   \frac{Dv}{Dt} + \left ( f + \frac{v}{r} \right ) u +  \frac{\nu}{r} \frac{\partial p}{\partial \phi} = D_v,
   \label{azimuthal} 
\end{equation}
\begin{equation}
    \frac{Dw}{Dt} + \nu \frac{\partial p}{\partial z} = -g + D_w ,
    \label{vertical}
\end{equation}
\begin{equation}
    \frac{D\eta}{Dt} = \frac{\dot{q}}{T} ,
    \label{thermodynamics}
\end{equation}
\begin{equation}
    \frac{\partial \rho}{\partial t} + \frac{1}{r} \frac{\partial (\rho r u)}{\partial r} +
    \frac{1}{r} \frac{\partial (\rho v)}{\partial \phi}  + 
    \frac{\partial (\rho w)}{\partial z} = 0 ,
    \label{continuity}
\end{equation}
\begin{equation}
     \frac{D}{Dt} = \frac{\partial}{\partial t} + u \frac{\partial}{\partial r}
     + \frac{v}{r} \frac{\partial}{\partial \phi} + w \frac{\partial}{\partial z},
     \label{lagrangian_derivative}
\end{equation}
where $(u,v,w)$ is the velocity field, $p$ is pressure, $\rho$ is the density, $\nu = 1/\rho$ is the specific volume, $\eta$ is the specific entropy,
$g$ is the acceleration of gravity, $r$ is the radial coordinate increasing outward, $z$ is height increasing upward. 
The terms $D_i$, $i=u,v,w$ denote dissipative terms for momentum, while $\dot{q}$ denotes diabatic heating. The thermodynamic equation of state is assumed in the form $\rho = \rho(\eta,p)$ or $\nu = \nu(\eta,p)$. 
For the developments that follow, it is useful to rewrite Eq. (\ref{azimuthal}) for the azimuthal motion in terms of the specific angular momentum $M = r v + f r^2/2$ as
\begin{equation}
    \frac{DM}{Dt} = r D_v - \frac{1}{\rho} \frac{\partial p}{\partial \phi} .
    \label{angular_momentum}
\end{equation}
As expected, $M$ is materially conserved for purely axisymmetric motions ($\partial p/\partial \phi=0)$ in the absence of the dissipative term $D_v$. 
\textcolor{violet}{From now on, only the axisymmetric case is considered.}
The following relations expressing various quantities in terms of $M$ will prove useful:
\begin{equation}
       v = \frac{M}{r} - \frac{fr}{2},
\end{equation}
\begin{equation}
    \frac{v^2}{2} = \frac{M^2}{2 r^2} + \frac{f^2 r^2}{8} 
        - \frac{f M}{2} = \mu \chi + \frac{f^2}{16 \chi} - \frac{f\sqrt{\mu}}{2} ,
        \label{v2intermsofM}
\end{equation}
\begin{equation}
    \left ( f + \frac{v}{r} \right ) v = \frac{M^2}{r^3} - \frac{f^2 r}{4}  = - \left  ( \mu - \frac{f^2}{16 \chi^2} \right )
    \frac{\partial \chi}{\partial r} ,
\end{equation}
where we have defined $\chi = 1/(2 r^2)$ and $\mu = M^2$, similarly to \citet{Andrews2006}. Note that 
(\ref{v2intermsofM}) assumes $M>0$ in order to write $M=\sqrt{\mu}$. Other quantities of importance in the following discussions are the vorticity 
\begin{equation}
      {\bf \xi} = - \frac{\partial v}{\partial z} \hat{\bf r} + \left ( \frac{\partial u}{\partial z} - \frac{\partial w}{\partial r}
      \right ) \hat{\bf \phi} + \frac{1}{r} \frac{\partial ( r v)}{\partial r} \hat{\bf z} 
\end{equation}
and potential vorticity $Q = ({\bf \xi} + f  \hat{\bf z})\cdot \nabla \eta/\rho$. It is useful to remark that $M$ and $Q$ are linked through the relation:
\begin{equation}
      Q = \frac{1}{\rho r} \frac{\partial (M,\eta)}{\partial (r,z)} .  
\end{equation}
Potential vorticity is thus proportional to the Jacobian of the coordinate transformation allowing one to map the physical space $(r,z)$ to the space $(M,\eta)$ of the materially conserved quantities for axisymmetric motions. As discussed later on, the stability of axisymmetric compressible vortex motions depends crucially on $Q$ being single-signed over the domain considered. 

\section{Available energetics}

\label{available_energetics} 

\subsection{Reference states}

\begin{figure}
    \centering
    \includegraphics[width=12cm]{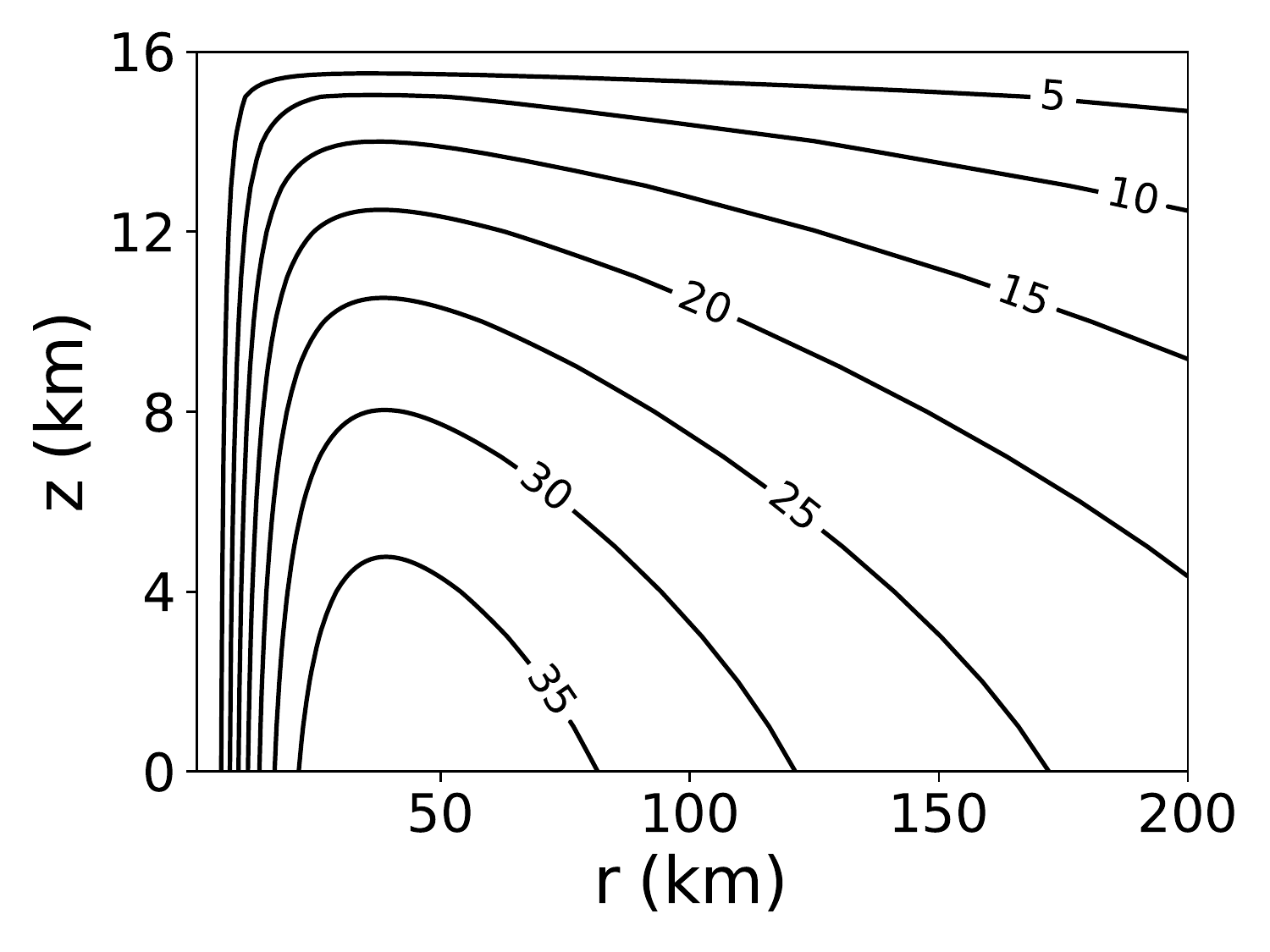}
    \caption{Azimuthal wind speed $v_m$ of the analytical dry vortex from \citet{Smith2005} used as a reference state to illustrate the momentum-constrained available energy. Contour labels indicate speed in $\mathrm{m\cdot s^{-1}}$.}
    \label{fig:v_contours}
\end{figure}

Following \citet{Andrews2006}, we define the reference state pertaining to the construction of momentum-constrained available energy as an axisymmetric solution of the inviscid form of Eqs. (\ref{radial}--\ref{continuity}). For such a reference state, the azimuthal velocity $v_m(r,z)$, pressure $p_m(r,z)$ and density $\rho_m(r,z)$ are in gradient wind and hydrostatic balances:
\begin{equation}
       \frac{1}{\rho_m} \frac{\partial p_m}{\partial r} = \left ( f + \frac{v_m}{r} \right ) v_m , \qquad
       \frac{1}{\rho_m} \frac{\partial p_m}{\partial z} = - g .
       \label{gradient_wind_balance}
\end{equation}
The corresponding reference profiles $\eta_m(r,z)$ and $M_m(r,z)$ for the specific entropy and angular momentum may then be inferred from the equation of state for density $\rho_m(r,z) = \rho(\eta_m(r,z),p_m(r,z))$, and via the definition of angular momentum $M_m(r,z) = r v_m(r,z) + f r^2/2$. For illustrative purposes, Fig. \ref{fig:v_contours} shows a particular example of azimuthal wind speed associated with the analytical dry atmospheric vortex solution used by \citet{Smith2005}, whose details can be found in Appendix \ref{analytical_solution}. This analytical solution serves as the basis for all subsequent illustrations. 
\textcolor{violet}{In the TC literature, the partitioning of an axisymmetric vortex into a balanced and unbalanced component has tended to be based on a blend of ad-hoc and plausible physical considerations, one of the primary considerations being ease of practical implementation, some of the most important examples of constructions being \citet{Nolan2002} or \citet{Smith2006}. From a theoretical viewpoint, however, it is essential to consider a partitioning into balanced and unbalanced components that is a priori physically well-defined (preferably uniquely) and amenable to theoretical analysis. As far as we are aware, there is really only one physical situation for which is this is the case, namely when the unbalanced part can be regarded as an inviscid and adiabatic perturbation to a symmetrically and inertially stable balanced part. From a formal viewpoint, this leads one to regard the balanced part of the vortex as the state of minimum energy obtainable by means of an adiabatic and inviscid re-arrangement of mass, as argued by \citet{Cullen2015} and \citet{Methven2015}, similarly as for the construction of the background reference state in \citet{Lorenz1955}'s theory of available potential energy. How one might implement such a construction in practice is not really understood but as we show below, such an approach allows one to obtain definite theoretical results about the evolution of the balanced and unbalanced parts of the vortex.}

\textcolor{violet}{Mathematically, the idea that the unbalanced part of the vortex represents an adiabatic and inviscid perturbation to the balanced reference state is encapsulated in the following relations: }
\begin{equation}
        M_m(r_{\star},z_{\star}) = M, \qquad \eta_m(r_{\star},z_{\star}) = \eta ,
        \label{reference_state_system}
\end{equation}
\textcolor{violet}{where $r_{\star}$ and $z_{\star}$ denote the coordinates of a fluid parcel in the vortex reference state.} Eq. (\ref{reference_state_system}) generalises the level of neutral buoyancy (LNB) equation introduced in \citet{Tailleux2013b} \textcolor{violet}{and states that a fluid parcel in the reference state has the same entropy and angular momentum as in the actual state. As we show below, such relations are the key to most of the subsequent theoretical results. In the most general case, $M_m$ and $\eta_m$ should be regarded as time-dependent. For simplicity, this time-dependence is neglected in this section but explicitly accounted for in the next section, as this is essential for addressing the effects of sinks and sources of entropy and angular momentum.}
To simplify subsequent derivations, one can rewrite (\ref{gradient_wind_balance}) as
\begin{equation}
          \frac{1}{\rho_m} \nabla p_m = \left ( \frac{M_m^2}{r^3} - \frac{f^2 r}{4} \right ) \nabla r - \nabla \Phi = 
          - \left ( \mu_m  - \frac{f^2}{16 \chi^2} \right ) \nabla \chi - \nabla \Phi,
          \label{phi_differential}
\end{equation}
where $\Phi = g_0 z$ is the geopotential ($g_0$ is gravitational acceleration).

To clarify the nature of the links with standard APE theory, we also consider a notional state of rest defined to be a function of height only and in hydrostatic balance only, viz., 
 \begin{equation}
            \frac{1}{\rho_0} \nabla p_0 = - \nabla \Phi
 \end{equation}
 where $\rho_0(z) = \rho(\eta_0(z),p_0(z))$. In that case, the reference entropy profile is defined through
 $\rho_0(z) = \rho(\eta_0(z),p_0(z))$ and the reference angular momentum is a function of radius only:
 $M_0(r) = fr^2/2$. One may similarly define a reference position $(r_R,z_R)$ in such a reference state via
 the equations:
 \begin{equation}
           \eta_0(z_R) = \eta , \qquad  \frac{f r_R^2}{2} = M . 
 \end{equation}
 Eqs. (\ref{reference_state_system}) represent constraints similar to those underlying the Generalised Lagrangian Mean (GLM) theory of \citet{Andrews1978}, in which the zero mean assumption for the displacements is relaxed (see \citet{Gilbert2018} for a recent revisiting of GLM theory). \textcolor{blue}{The radius $r_R$ is referred to as the potential radius by \citet{Emanuel1995}.} 

\subsection{Available energetics}

Prior to discussing available energetics, one must start by identifying the appropriate form of total energy conservation pertaining to the system of equations considered. In the present case, the standard form of total energy is  ${\bf v}^2/2+e + \Phi = {\bf v}^2/2 + h + \Phi - p/\rho$, where $e$ is internal energy. Following the usual procedure, the latter can be shown to satisfy the following evolution equation: 
\begin{equation}
   \rho \frac{D}{Dt} \left ( \frac{{\bf v}^2}{2} + h + \Phi - \frac{p}{\rho} \right ) + 
   \nabla \cdot (p {\bf v}) = \rho {\bf v} \cdot {\bf D} + \rho \dot{q} .
   \label{total_energy}
\end{equation}
For the system of equations considered to be energetically consistent, the viscous and diabatic terms ${\bf D}$ and $\dot{q}$ must be such that the right-hand side of Eq. (\ref{total_energy}) is expressible as the divergence of some flux. Following \citet{Andrews1981, Andrews2006}, we introduce the following identity:
\begin{equation}
     \rho \frac{D}{Dt} \left ( \frac{p_m}{\rho} \right ) = \frac{\partial p_m}{\partial t} + 
     \nabla \cdot ( p_m {\bf v} ) ,
\end{equation}
which is valid for any arbitrary pressure field $p_m = p_m({\bf x},t)$, and add it to 
(\ref{total_energy}) to obtain the following alternative energy conservation equation:
%\begin{equation}
%      \rho \frac{D}{Dt}\left ( \frac{{\bf v}^2}{2} + h + \Phi + \frac{p_m-p}{\rho} \right ) 
%      + \nabla \cdot [ (p-p_m) {\bf v} ] = \rho {\bf v} \cdot {\bf D} + \rho \,\dot{q}  + \frac{\partial p_m}{\partial t} ,
%\end{equation}
\begin{equation}
      \rho \frac{D}{Dt} \left ( \frac{{\bf v}^2}{2} + \Pi_1 + h(\eta,p_m) + \Phi(z) \right )
     + \nabla \cdot [ (p-p_m){\bf v} ] = \rho {\bf v} \cdot {\bf D} + \rho \,\dot{q} + \frac{\partial p_m}{\partial t}, 
\end{equation}
where $\Pi_1 = h(\eta,p) - h(\eta,p_m) + (p_m-p)/\rho$ is a positive definite energy quantity usually referred to as Available Acoustic Energy (AAE), e.g. \citet{Andrews1981, Andrews2006}, \citet{Tailleux2018}. The positive definite character of $\Pi_1$ follows from the possibility to write it in the form
\begin{equation}
    \Pi_1 = \int_{p_m}^p \int_p^{p'} \nu_p(\eta,p'')\,{\rm d}p'' {\rm d}p' 
    = \int_{p_m}^{p} \int_{p'}^p \frac{{\rm d}p'' {\rm d}p'}{\rho^2 c_s^2} \approx \frac{(p-p_m)^2}{2 \rho_m^2 c_{s0}^2} ,
\end{equation}
where $c_s^2 = (\partial \rho/\partial p)^{-1}$ is the squared speed of sound
\citep{Tailleux2018}. The advantage of using a reference pressure field that also depends on the horizontal coordinates is that it reduces the magnitude of the pressure perturbation $p'$ as compared to standard APE theory, and hence reduces the contribution of AAE to the overall energy budget. Following \citet{Andrews2006} and \citet{Codoban2006}, we decompose the total kinetic energy as the sum of the kinetic energies of the secondary (toroidal) and primary (azimuthal) circulations:
\begin{equation}
       \frac{{\bf v}^2}{2} = \frac{{\bf u}_s^2}{2} + \frac{v^2}{2},
\end{equation}
where ${\bf u}_s = (u,0,w)$ is the velocity of the secondary circulation. Next, we define the vortex dynamic potential energy ${\cal V}$ as
%\begin{equation}
%    {\cal V} = \frac{v^2}{2} + h(\eta,p) + \Phi(z) + \frac{p_m - p}{\rho}
%\end{equation}
\begin{equation}
     {\cal V} = \frac{v^2}{2} + h(\eta,p_m) + \Phi(z) ,
\end{equation}
and define the vortex available energy $A_e$ as the difference between the values of ${\cal V}$ in the actual and reference states:
%\begin{equation}
%        A = {\cal V} - {\cal V}_{\star} = 
%        \frac{v^2}{2} - \frac{v_{\star}^2}{2} + h(\eta,p) - h(\eta,p_{\star}) + \Phi(z) - \Phi(z_{\star}) 
%        +  \frac{p_m-p}{\rho}
%\end{equation}
\begin{equation}
         A_e = {\cal V} - {\cal V}_{\star} = \frac{v^2}{2} - \frac{v_{\star}^2}{2}
          + h(\eta,p_m) - h(\eta,p_{\star}) + \Phi(z) - \Phi(z_{\star}), 
          \label{available_energy_definition}
\end{equation}
where $p_{\star} = p_m(r_{\star},z_{\star},t)$ and $\chi_{\star} = \chi(r_{\star})$. \textcolor{violet}{Moreover, note that by construction 
\begin{equation}
     v_{\star} = \frac{M}{r_{\star}} - \frac{f r_{\star}}{2} = \frac{M_m(r_{\star},z_{\star})}{r_{\star}} 
     - \frac{f r_{\star}}{2} = 
     v_m (r_{\star},z_{\star}) . 
\end{equation}
}
Note that if it were not for the presence of the kinetic energy term $v^2/2-v_{\star}^2/2$, $A_e$ would be nearly identical to the eddy APE term introduced by \citet{Tailleux2018}, which motivates us to regard $A_e$ as a particular form of eddy energy. To discuss the properties of $A_e$, it is more advantageous to express $v^2/2$ and $v_{\star}^2/2$ in terms of $\mu$ by using (\ref{v2intermsofM}) as follows:
\begin{equation}
      A_e = \mu (\chi - \chi_{\star})  + \frac{f^2}{16} \left ( \frac{1}{\chi} - \frac{1}{\chi_{\star}} \right ) 
       + h(\eta,p_m) - h(\eta,p_{\star}) + \Phi(z) - \Phi(z_{\star}). 
       \label{ae_definition}
\end{equation}
Eq. (\ref{ae_definition}) is the starting point for the subsequent derivations.

\subsection{Interpretation of $A_e$ in terms of the work of a generalised buoyancy force} 
As can be seen from (\ref{ae_definition}), a key property of $A_e$ is that it is a function of the actual and reference positions only at fixed $\eta$ and $\mu$ (or $M$). As a result, it is possible to write $A_e$ as the path integral
\begin{equation}
      A_e = - \int_{{\bf x}_{\star}}^{{\bf x}} 
         {\bf b}_e(\mu,\eta,{\bf x}',t) \cdot {\rm d}{\bf x}' , 
         \label{path_integral_ae}
\end{equation}
thus allowing $A_e$ to be interpreted as the work against the generalised buoyancy force ${\bf b}_e$ defined by
\begin{equation}
    {\bf b}_e =  - \nabla A_e = \underbrace{\left ( 
    \nu_m - \nu_h \right ) \nabla p_m}_{{\bf b}_e^T} + 
    \underbrace{(\mu_m-\mu) \nabla \chi}_{{\bf b}_e^M} ,
    \label{generalised_buoyancy}
\end{equation}
where the derivatives are taken by holding $\eta$ and $\mu$ constant, with ${\bf x}=(r,z)$ and ${\bf x}_{\star} = (r_{\star},z_{\star})$. This result is interesting and important, because it generalises to $A_e$ the possibility first established by \citet{Andrews1981} to interpret the APE density as the work needed to move a fluid parcel from its notional equilibrium position ${\bf x}_{\star}$ in the reference state to its position ${\bf x}$ in the actual state. Whether such a possibility also pertained to momentum-constrained available energy had remained unclear so far, as neither \citet{Codoban2006} nor \citet{Andrews2006} had discussed it. 

In the following, we regard the part ${\bf b}_e^T$that is proportional to the specific volume anomaly as the thermodynamic component and the part ${\bf b}_e^M$ that is proportional to the squared angular momentum anomaly as the mechanical component of the generalised buoyancy force ${\bf b}_e$. The thermodynamic force ${\bf b}_e^T$ has been discussed before by \citet{Smith2005}, who sought to clarify the role played by buoyancy in tropical cyclones. In their paper, they write such a force in the form
\begin{equation}
       {\bf b}_e^T = \left ( 1 - \frac{\rho_m}{\rho_h} \right ) \frac{1}{\rho_m} \nabla p_m = \left ( 1-\frac{\rho_m}{\rho_h} \right ) {\bf g}_e,
       \label{generalised_buoyancy_thermo}
\end{equation}
where ${\bf g}_e$ is a generalised acceleration defined by \citet{Smith2005} as
\begin{equation}
      {\bf g}_e = \left ( \frac{v_m^2}{r} + f v_m, -g \right ) = \frac{1}{\rho_m} \nabla p_m .
\end{equation}
The mechanical force ${\bf b}_e^M$ has also been discussed before, for instance in relation to centrifugal waves, e.g. \citet{Markowski2010}.

\subsection{Partitioning of $A_e$ into mechanical and thermodynamic components}

The available energy $A_e$ is a two-dimensional function of the radial and vertical displacements $\delta r = r - r_{\star}$ and $\delta z = z-z_{\star}$. For small enough displacements, $A_e$ reduces to a quadratic function $A_e \approx 1/2 {\delta {\bf x}}^T H_A \delta {\bf x}$, where $H_A$ is the Hessian matrix of $A_e$'s second derivatives. In that case, the most natural approach to establish the positive definite character of $A_e$ is to compute the (real) eigenvalues of $H_A$ and determine whether they are both positive. Since ${\bf b}_e = -\nabla A_e \approx - H_A \delta {\bf x}$, the eigenvectors of $H_A$ can be interpreted as the directions along which the restoring buoyancy force ${\bf b}_e$ aligns perfectly with the displacement $\delta {\bf x}$. The eigenvalues thus represent the squared frequency of the natural oscillation taking place in the eigendirections. In the system of coordinates $(x_1,x_2)$ defined by the eigenvectors, the available energy $A_e$ can be written as the sum of two quadratic terms $A_e \approx \lambda_1 x_1^2/2 + \lambda_2 x_2^2/2$ that most clearly links its sign positive definite character to the sign of the eigenvalues $\lambda_1$ and $\lambda_2$. Such an approach is not available, however, for finite-amplitude displacements. In that case, previous authors have shown that $A_e$ can still be decomposed as the sum of two terms whose sign-definiteness can be linked to the sign of the gradients of $M_m$ and $\eta_m$. Such a decomposition is not unique in general, however, and different authors have proposed different ones. 

Here we propose yet another decomposition $A_e = \Pi_k + \Pi_e$, which differs from previous ones in that it attempts to more cleanly separate the kinetic and potential energy parts of $A_e$. To that end, we exploit the fact that $A_e$ can be expressed as a path integral. Indeed, such a path can be broken into two components by introducing some yet-to-be-identified intermediate point ${\bf x}_{\mu} = (r_{\mu},z_{\mu})$, thus allowing one to associate $\Pi_e$ and $\Pi_k$ with one of the integration sub-paths as follows:
\begin{equation}
     A_e = \underbrace{-\int_{{\bf x}_{\star}}^{{\bf x}_{\mu}} 
     {\bf b}_e(M,\eta,{\bf x}',t)\cdot {\rm d}{\bf x}'}_{\Pi_e} \underbrace{- \int_{{\bf x}_{\mu}}^{{\bf x}} {\bf b}_e(M,\eta,{\bf x}',t)\cdot {\rm d}{\bf x}'}_{\Pi_k} .
\end{equation}
The same intermediate point ${\bf x}_{\mu}$ can be similarly used to partition the exact expression (\ref{ae_definition}) for $A_e$, thus yielding the following explicit expressions for $\Pi_e$ and $\Pi_k$:
\begin{equation}
    \Pi_e = 
    h(\eta,p_\mu) - h(\eta,p_{\star}) + \Phi(z_{\mu}) - \Phi(z_{\star}) 
      + \mu (\chi_{\mu}-\chi_{\star}) + \frac{f^2}{16} \left ( \frac{1}{\chi_{\mu}} - \frac{1}{\chi_{\star}} \right ) ,
      \label{pie_analytical}
\end{equation}
\begin{equation}
     \Pi_k = h(\eta,p_m) - h(\eta,p_\mu) 
     + \Phi(z) - \Phi(z_\mu) + \mu (\chi-\chi_\mu) + \frac{f^2}{16} \left ( \frac{1}{\chi}-\frac{1}{\chi_{\mu}} 
     \right ) ,
     \label{pik_analytical} 
\end{equation}
where $p_{\mu}$ is shorthand for $p_m(r_{\mu},z_{\mu})$. In Eq. (\ref{pik_analytical}), the terms involving the specific enthalpy are clearly of a thermodynamic nature. In order for $\Pi_k$ to be purely mechanical in nature, these need to be removed. This is possible only if $(r_{\mu},z_{\mu})$ is chosen so that $p_{\mu} = p_m$. To further constrain ${\bf x}_{\mu}$, we further impose that $\Pi_e$ and $\Pi_k$ are only contributed to by the thermodynamic and mechanical components of the generalised buoyancy force respectively; mathematically:
\begin{equation}
      \Pi_e = - \int_{{\bf x}_\star}^{{\bf x}_{\mu}} 
      {\bf b}_e^T\cdot {\rm d}{\bf x}'  = 
      \int_{{\bf x}_{\star}}^{{\bf x}} (\nu_h - \nu_m)\nabla p_m \cdot {\rm d}{\bf x}'
      \label{pie_expression}
\end{equation}
\begin{equation}
      \Pi_k = - \int_{{\bf x}_{\mu}}^{{\bf x}} {\bf b}_e^M\cdot {\rm d}{\bf x}' = \int_{{\bf x}_{\star}}^{{\bf x}} (\mu-\mu_m)\nabla \chi \cdot {\rm d}{\bf x}' 
      .
      \label{pik_expression}
\end{equation}
For such expressions to hold, the work done by ${\bf b}_{e}^M$ and ${\bf b}_e^T$ must vanish on the first and second legs of the overall integration path respectively:
\begin{equation}
    \int_{{\bf x}_{\mu}}^{{\bf x}} (\nu_h - \nu_m)\nabla p_m \cdot {\rm d}{\bf x}'  = 0 , \qquad \int_{{\bf x}_{\star}}^{{\bf x}_{\mu}} (\mu - \mu_m)\nabla \chi \cdot {\rm d}{\bf x}' = 0 .
    \label{conditions_to_fulfill} 
\end{equation}
The most obvious way to fulfill (\ref{conditions_to_fulfill}) is by imposing $\mu_m = \mu$ on the first leg joining ${\bf x}_{\star}$ to ${\bf x}_{\mu}$, while imposing to the second leg joining ${\bf x}_{\mu}$ to ${\bf x}$ that it follows an isobaric surface $\nabla p_m \cdot {\rm d}{\bf x}' = 0$. As a result, the intermediate point ${\bf x}_{\mu}= (r_{\mu},z_{\mu})$ must lie at the intersection of the surface of constant angular momentum $\mu_m = \mu$ and isobaric surface $p_m = p_m(r,z)$. Its coordinates must therefore be solutions of
\begin{equation}
           \mu_m (r_{\mu},z_{\mu}) = \mu, \qquad p_m (r_{\mu},z_{\mu}) = p_m(r,z) .
           \label{intersection_point}
\end{equation}
Such a construction and the two different integration paths are illustrated in Fig. \ref{fig:lifting_illustration} for the analytical vortex solution detailed in Appendix \ref{analytical_solution}. Moreover, Fig. \ref{fig:ae_perturb} illustrates the superiority, at least visually, of the $(\mu_m,p_m)$ representation over the $(M_m,\eta_m)$ representation to achieve what looks like a near orthogonal finite-amplitude decomposition of available energy.

\begin{figure}
    \centering
    \includegraphics[width=0.9\textwidth]{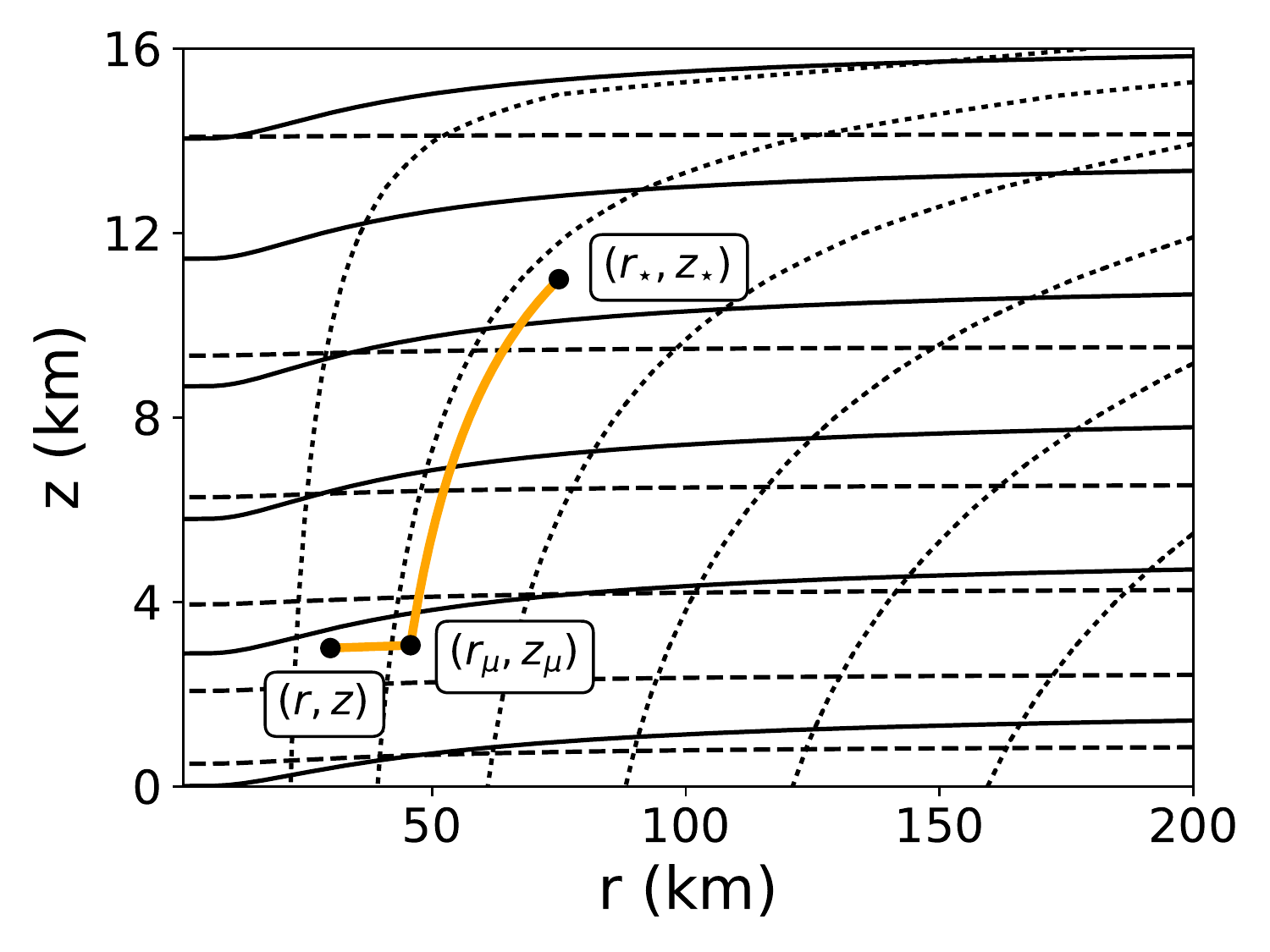}
    \caption{Illustration of a particular pathway linking a fluid parcel reference position $(r_{\star},z_{\star})$ to its actual position $(r,z)$ via the intermediate point $(r_{\mu},z_{\mu})$ for the analytical vortex state described in Appendix \ref{analytical_solution}. The first leg of the path linking ${\bf x}_{\star}$ to ${\bf x}_{\mu}$ follows a surface of constant angular momentum (denoted by dotted lines). The second leg linking ${\bf x}_{\mu}$ to ${\bf x}$ follows an isobaric surface (denoted by dashed lines). The thick full lines denote isentropic surfaces, which illustrate the warm core character of such a cyclonic vortex. }
    \label{fig:lifting_illustration}
\end{figure}

\subsection{Sign definiteness of the eddy available potential energy $\Pi_e$}

The key role played by isobaric and constant angular momentum surfaces for simplifying the partitioning of $A_e$ motivates us to work in $(\mu_m,p_m)$ coordinates. Since $\mu_m = (r^3/\rho_m)\partial p_m/\partial r + f^2 r^4/4$, we have
\begin{equation}
   J_{\mu p} = \frac{\partial (\mu_m,p_m)}{\partial (r,z)} =  -\rho_m g f^2 r^3 
   + \frac{\partial}{\partial r} \left ( \frac{r^3}{\rho_m} 
   \frac{\partial p_m}{\partial r} \right ) \frac{\partial p_m}{\partial z}
    - \frac{\partial}{\partial z} \left ( \frac{r^3}{\rho_m} \frac{\partial p_m}{\partial r} \right ) \frac{\partial p_m}{\partial r} .
    \label{jmup}
\end{equation} 
Close to a state of rest, $J_{\mu p} \approx -\rho_m g f^2 r^3 <0$, in which case the coordinate transformation is invertible and well-defined, except at the origin $r=0$. In the following, we assume that this remains the case for a non-resting reference vortex state in gradient wind balance, as can be seen to be the case in Fig. \ref{fig:lifting_illustration} for our example analytical vortex. In the following, a tilde is used to denote functions of $(r,z)$ in their $(\mu_m,p_m)$ representation; for instance, $\nu_m(r,z) = \tilde{\nu}_m(\mu_m,p_m)$ and $\chi(r) = \tilde{\chi}(\mu_m,p_m)$. 

%\begin{figure}
%    \centering
%    %\includegraphics{}
%    \caption{Ratio of the Jacobian for the particular vortex solution over the %Jacobian value for a state of rest.}
%    \label{fig:jacobian}
%\end{figure}

Using the fact that $\nabla p_m \cdot {\rm d}{\bf x}' = {\rm d}p'$, where $p'=p_m(r',z')$, it is easily seen that Eq. (\ref{pie_expression}) defining $\Pi_e$ may be rewritten as the following one-dimensional integral in pressure: 
\begin{equation}
    \Pi_e = \int_{p_{\star}}^{p_m} \left [ \nu(\eta,p') - \tilde{\nu}_m(\mu, p') \right ] \,{\rm d}p'  = \int_{p_{\star}}^{p_m} \left [ \nu(\eta,p') - \nu(\tilde{\eta}_m(\mu,p'),p') \right ] \,{\rm d}p' ,
     \label{pie_expression_ter}
\end{equation}
where we also used the fact that $\nu_h$ in Eq. (\ref{pie_expression}) refers to $\nu(\eta,p')$, that $p_m ({\bf x}_{\star}) = p_{\star}$, that $p_m = p_m ({\bf x})$, and that $\nu_m(r',z') = \tilde{\nu}_m(\mu,p') = \nu(\tilde{\eta}_m(\mu,p'),p')$ along the surface of constant angular momentum $\mu_m = \mu$. Physically, Eq. (\ref{pie_expression_ter}) can be recognised as being similar to the conventional APE density (compare with Eq. (2.18) of \citet{Tailleux2018}), for a definition of buoyancy defined relative to the horizontally-varying reference specific volume $\tilde{\nu}_m(\mu,p)$ evaluated along a constant angular momentum surface. As a result, $\Pi_e$ represents a ``slantwise'' APE density, by analogy with the concept of slantwise convective available potential energy (SCAPE) used in discussions of conditional symmetric instability \citep{Bennetts1979,Emanuel1983b,Emanuel1983a}. To establish the positive definite character of $\Pi_e$, note that (\ref{pie_expression_ter}) may be rewritten as
$$
     \Pi_e = \int_{p_{\star}}^{p_m} \int_{\tilde{\eta}_m(\mu,p')}^{\eta} \frac{\partial \nu}{\partial \eta}(\eta',p')\,{\rm d}\eta' {\rm d}p' 
     = \frac{\partial \nu}{\partial \eta}(\eta_i,p_i) 
     \int_{p_{\star}}^{p_m} \int_{\tilde{\eta}_m(\mu,p')}^{\tilde{\eta}_m(\mu,p_{\star})} {\rm d}\eta' {\rm d}p' 
$$
\begin{equation} 
     = \frac{\partial \nu}{\partial \eta}(\eta_i,p_i) 
     \int_{p_{\star}}^{p_m} \int_{p'}^{p_{\star}}  \frac{\partial \tilde{\eta}_m}{\partial p}(\mu,p'')\,{\rm d}p'' {\rm d}p' ,
     \label{pie_expression_bis}
\end{equation}
where we have used the mean value theorem to take the adiabatic lapse rate $\partial \nu/\partial \eta = \Gamma = \alpha T/(\rho c_p)$ out of the integral ($\alpha$ is the isobaric thermal expansion and $c_p$ is the isobaric specific heat capacity), where $(\eta_i,p_i)$ represent some intermediate values of entropy and pressure, and used the fact that $\eta = \tilde{\eta}_m(\mu,p_{\star})$ by definition. If the adiabatic lapse rate $\Gamma$ is positive, as is normally the case, Eq. (\ref{pie_expression_bis}) shows that a sufficient condition for $\Pi_e$ to be positive definite is
\begin{equation}
     \frac{\partial \tilde{\eta}_m}{\partial p}(\mu,p'') < 0 , 
     \label{stability_pie} 
\end{equation}
regardless of $p''$; this states that the specific entropy should increase with height (decrease with pressure) along surfaces of constant angular momentum, as expected. The special case where 
\begin{equation}
     \frac{\partial \eta_m}{\partial z}(r,z) > 0, 
     \qquad \frac{\partial \tilde{\eta}_m}{\partial p}(\mu,p'') > 0, 
\end{equation}
would correspond to the so-called conditional symmetric instability (CSI), whereby the entropy profile is stable to upright vertical displacements but not to slantwise displacements. For small amplitude perturbations, a Taylor series expansion shows that (\ref{pie_expression_bis}) approximates to
\begin{equation}
        \Pi_e \approx - \Gamma_i \frac{\partial \tilde{\eta}_m}{\partial p}(\mu,p_{\star}) \frac{(p_m-p_{\star})^2}{2} 
        \label{small_amplitude_pie} 
\end{equation}
where $\Gamma_i$ is shorthand for $\partial \nu/\partial p(\eta_i,p_i)$. Note that this expression is essentially the same as the classical small-amplitude expression $N^2 \delta z^2/2$ for the conventional APE density in terms of an appropriate squared buoyancy frequency, where $\delta z$ is the vertical displacement from the reference height. Note that in Eq. (\ref{pie_expression_bis}), we could equally have regarded pressure as a function of entropy to obtain a small amplitude approximation proportional to the squared entropy anomaly $(\tilde{\eta}_m(\mu,p)-\tilde{\eta}_m(\mu,p_{\star}))^2/2$ instead if desired.

\begin{figure}
    \centering
    \begin{subfigure}{0.49\textwidth}
        \includegraphics[width=\textwidth]{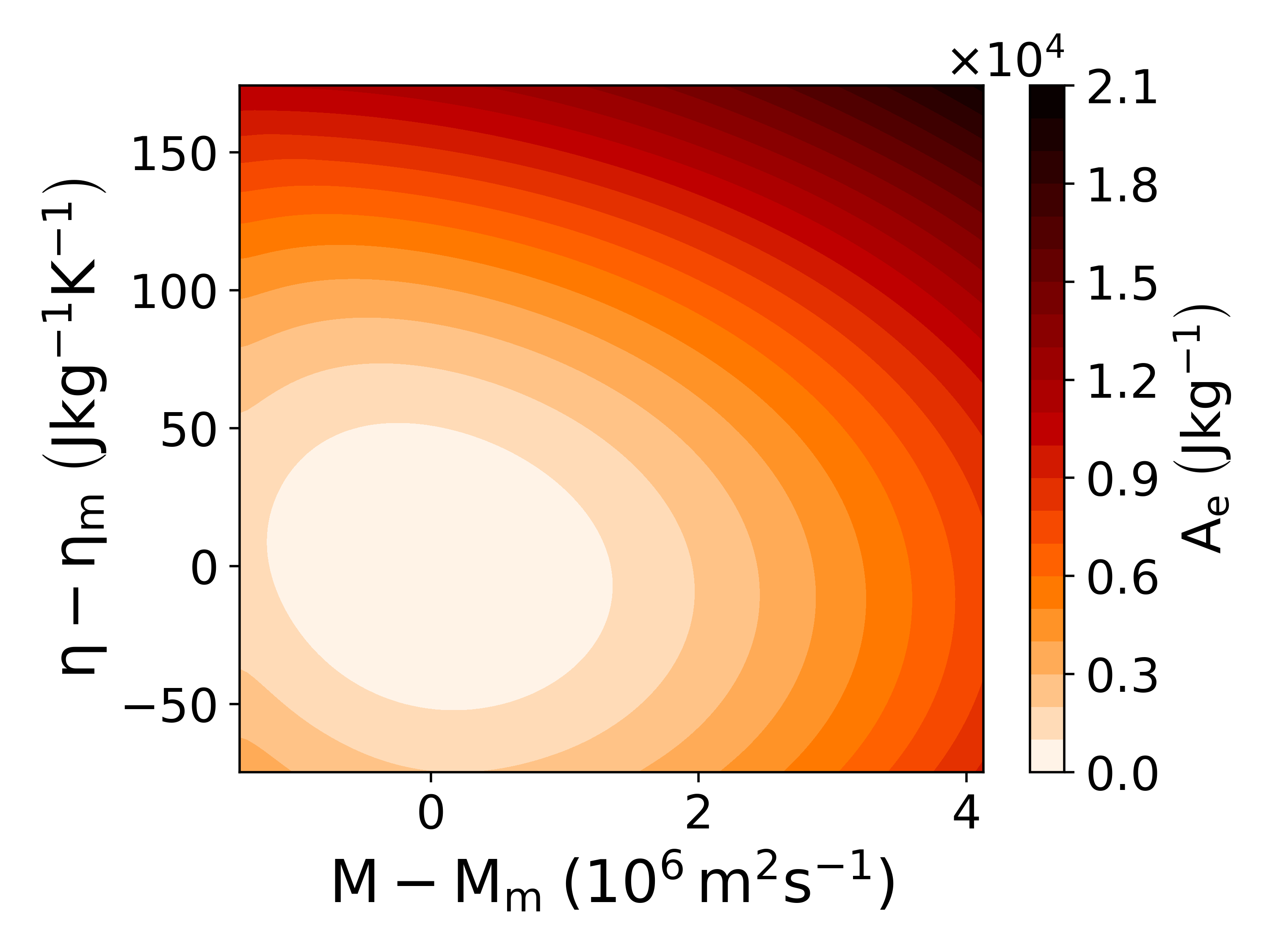}
        \caption{}
        \label{fig:ae_perturb_M_eta}
    \end{subfigure}
    \begin{subfigure}{0.49\textwidth}
        \includegraphics[width=\textwidth]{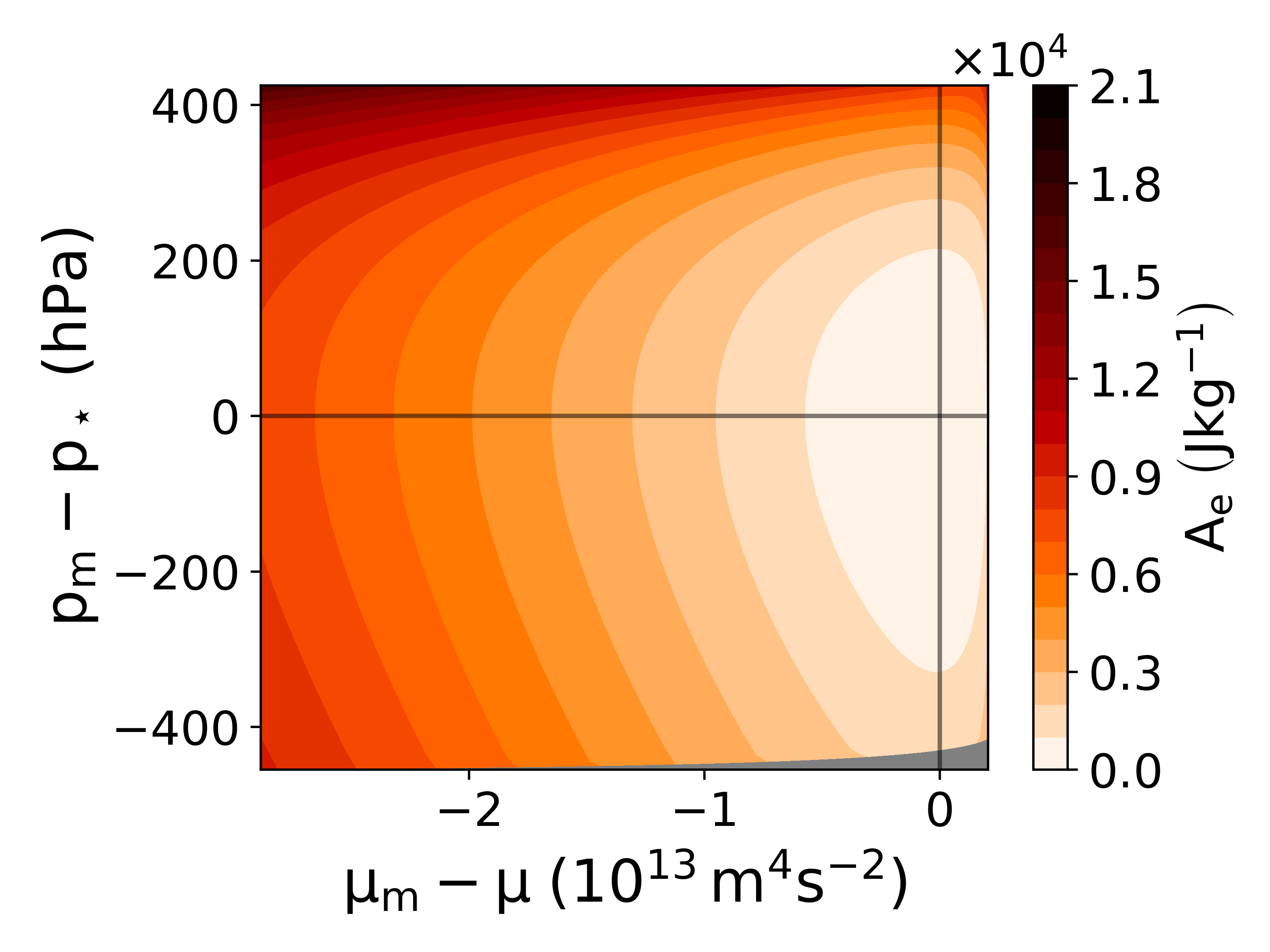}
        \caption{}
        \label{fig:ae_perturb_mu_p}
    \end{subfigure}
    \caption{Available energy $A_e$ of a perturbed dry air parcel at $r=\SI{40}{\kilo\metre}$, $z=\SI{5}{\kilo\metre}$, in terms of (\subref{fig:ae_perturb_M_eta}) $M$ and $\eta$ perturbations and (\subref{fig:ae_perturb_mu_p}) $\mu$ and $p^\star$ perturbations. The grey lines in (\subref{fig:ae_perturb_mu_p}) indicate the horizontal and vertical axes along which $\Pi_k$ and $\Pi_e$ change respectively, and the grey shading covers points in the space that are not sampled by the chosen perturbations of $M$ and $\eta$.}
\label{fig:ae_perturb}
\end{figure}

\subsection{Sign definiteness of the mechanical eddy energy $\Pi_k$} 

We now turn to the problem of establishing the conditions for $\Pi_k$ to be positive definite. Eqs. (\ref{pik_analytical}), (\ref{pik_expression}) and (\ref{conditions_to_fulfill}) show that $\Pi_k$ may be equivalently written as:
\begin{equation}
    \Pi_k = \Phi(z) - \Phi(z_{\mu}) + \mu (\chi - \chi_{\mu}) 
    + \frac{f^2}{16} \left ( \frac{1}{\chi} - \frac{1}{\chi_{\mu}} \right ) 
    =  \int_{{\bf x}_{\mu}}^{{\bf x}} (\mu-\mu_m)\nabla \chi \cdot {\rm d}{\bf x}' . 
\end{equation}
As for $\Pi_e$, we find it useful to keep working in $(\mu_m,p_m)$ coordinates. To that end, we use the mathematical identity $(\mu-\mu_m) \nabla \chi = \nabla [ (\mu-\mu_m) \chi ]  + \chi \nabla \mu_m$ (recall that $\mu$ is treated like a constant in such calculations), and the fact that by construction $p_\mu = p_m$, to rewrite $\Pi_k$ in the following equivalent ways: 
\begin{equation}
   \Pi_k =  (\mu - \mu_m ) \chi  + \int_{{\bf x}_{\mu}}^{{\bf x}} \tilde{\chi}(\mu_m,p_m) \nabla \mu_m \cdot {\rm d}{\bf x}'
%   = (\mu-\mu_m) \chi + \int_{\mu}^{\mu_m} \tilde{\chi}(\mu',p_m) {\rm d}\mu'
   = \int_{\mu}^{\mu_m} \left [ \tilde{\chi}(\mu',p_m)  - \tilde{\chi}(\mu_m,p_m) \right ]\,{\rm d}\mu' .
   \label{pik_expression_2} 
\end{equation}
Eq. (\ref{pik_expression_3}) is very similar to the expression Eq. (\ref{pie_expression}) for $\Pi_e$. It can be similarly expressed as a double integral,
\begin{equation}
    \Pi_k = \int_{\mu}^{\mu_m} \int_{\mu_m}^{\mu'} 
     \frac{\partial \tilde{\chi}}{\partial \mu} (\mu'',p_m) {\rm d}\mu'' {\rm d}\mu' ,
     \label{pik_expression_3} 
\end{equation}
which makes it clear that a sufficient condition for $\Pi_k$ to be positive definite is
\begin{equation}
        \frac{\partial \tilde{\chi}}{\partial \mu} ( \mu'', p_m) < 0 .
        \label{pik_positive}
\end{equation}
Physically, Eq. (\ref{pik_positive}) corresponds to the condition that the reference squared angular momentum distribution increase with radius along isobaric surfaces. The violation of this condition corresponds to centrifugal instability, e.g. \citet{Drazin1981}. It is useful to remark that the partial derivative $\partial \tilde{\chi}/\partial \mu$ can be expressed in terms of the Jacobian $J_{\mu p}$ of the coordinate transformation from $(r,z)$ to $(\mu_m,p_m)$ coordinates as
\begin{equation}
       \frac{\partial \tilde{\chi}}{\partial \mu} = \frac{\partial (\tilde{\chi},p_m)}{\partial (r,z)} 
       \left ( \frac{\partial (\mu_m,p_m)}{\partial (r,z)} \right )^{-1} 
        = \frac{\rho_m g}{r^3} 
       \left ( \frac{\partial (\mu_m,p_m)}{\partial (r,z)} \right )^{-1} ,
\end{equation}
which shows that the condition (\ref{pik_positive}) is actually equivalent to the condition $J_{\mu p}<0$, and hence to the requirement that the coordinates transformation from $(\mu_m,p_m)$ to $(r,z)$ be invertible. 

\begin{figure}
    \centering
    \includegraphics[width=10cm]{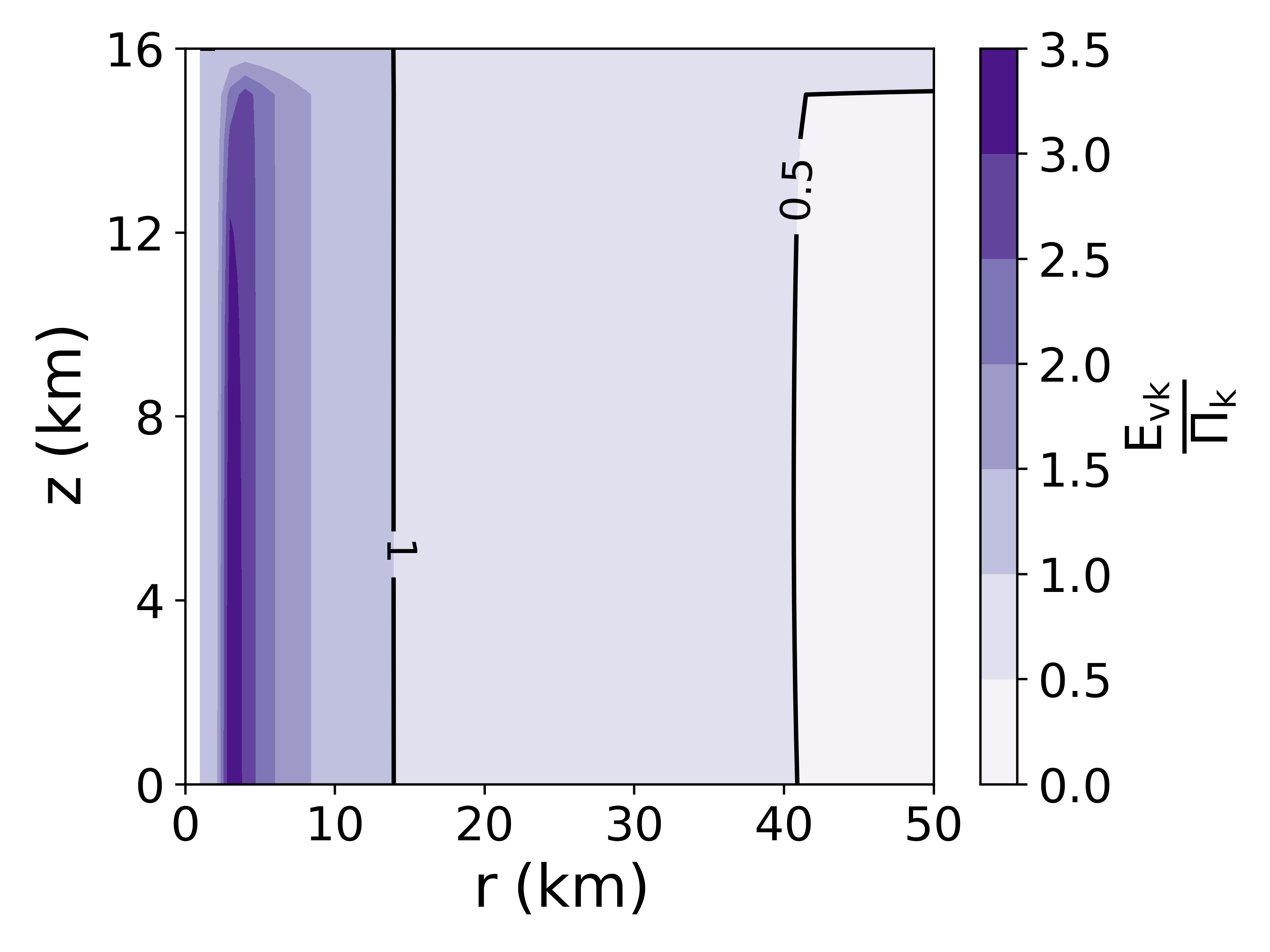}
    \caption{Estimate of the ratio $1/K_i$ as defined by Eq. (\ref{ratio_ki}) for the analytical vortex state described in Appendix \ref{analytical_solution}, assuming $M\approx M_m$.} 
    \label{fig:ratio_ki}
\end{figure}

By using the mean value theorem, it is possible to write
\begin{equation}
       \Pi_k = - \frac{\partial \tilde{\chi}}{\partial \mu}(\mu_i,p_m) 
       \frac{(\mu-\mu_m)^2}{2} 
\end{equation}
where $\mu_i$ is some intermediate value of $\mu$ within the interval $[{\rm min}(\mu,\mu_m),{\rm max}(\mu,\mu_m)]$. On the other hand, the Eulerian eddy kinetic energy $E_{vk} = (v-v_m)^2/2$ may be written as
\begin{equation}
    E_{vk} = \frac{(v-v_m)^2}{2} = \frac{1}{r^2(M_m+M)^2} \frac{(\mu_m-\mu)^2}{2} 
\end{equation}
by using the relations $v=M/r-fr/2$, $v_m = M_m/r - fr/2$, $(v-v_m)^2 = (M-M_m)^2/r^2$ and $(\mu_m-\mu)^2 = (M_m+M)^2(M-M_m)^2$. The two quantities are therefore proportional to each other, i.e., $\Pi_k = K_i E_{vk}$, with the proportionality factor
\begin{equation}
       K_i = - r^2(M_m+M)^2 \frac{\partial \tilde{\chi}}{\partial \mu}(\mu_i,p_m) . 
       \label{ratio_ki}
\end{equation}
Near a state of rest, $M_m \approx M \approx f r^2/2$, then 
$\partial \tilde{\chi}/\partial \mu \approx - 1/(f^2 r^6)$, so that $K_i \approx 1$, and the two quantities are equivalent. In general, however, $K_i \ne 1$ but we were not able to develop a mathematical theory for its value. For illustrative purposes, Fig. \ref{fig:ratio_ki} shows a particular estimate of the ratio $1/K_i$ for the particular example of the analytical reference vortex state described in Appendix \ref{analytical_solution}, where the approximation $\mu_i \approx \mu_m$ was used.

\subsection{Mean ``system'' energies}

\textcolor{violet}{The balanced reference vortex state possesses a conventional APE and KE relative to the horizontally uniform background reference hydrostatic and density fields $p_0(z,t)$ and $\rho_0(z,t)$ characteristic of the environment in which it is embedded. Based on existing local APE theory \citep{Tailleux2018}, the most natural definition of `mean' APE density, as confirmed below, is
\begin{equation}
        \Pi_m = h(\eta,p_{\star}) - h(\eta,p_R) + \Phi(z_{\star}) - \Phi(z_R)
    = \int_{p_R}^{p_{\star}} \nu(\eta,p') \,{\rm d}p'  - \int_{p_R}^{p_0(z_{\star},t)} \nu_0(p',t) \,{\rm d}p' ,
        \label{mean_APE_density} 
\end{equation}
where $z_R$ is the equilibrium position of the fluid parcel in the background reference state and is defined as the solution of the level of neutral buoyancy equation $\nu(\eta,p_R) = \nu_0(p_R,t)$, where $p_R = p_0(z_R,t)$. The quantity $\Pi_m$ is similar, but not identical, to the APE density denoted by $\Pi_2$ in \citet{Tailleux2018}. The difference arises because $p_{\star} = p_m(r_{\star},z_{\star},t)$ is in general different from $p_0(z_{\star},t)$. In the next section, such an approach is justified by showing that $\Pi_m$ is the main energy reservoir `feeding' the balanced vortex kinetic energy. In addition, it is also useful to define the locally-defined background potential energy as
\begin{equation}
     {\cal B}_R = h(\eta,p_R) + \Phi(z_R) .
     \label{background_potential_energy} 
\end{equation}
Using the fact that by construction,
\begin{equation}
     \frac{1}{\rho_m} \frac{\partial p_m}{\partial r} 
      = \frac{M_m^2}{r^3} - \frac{f^2 r}{4}, \qquad
      \frac{1}{\rho_m} \frac{\partial p_m}{\partial z} = - g , 
\end{equation}
\begin{equation}
     \frac{1}{\rho_0} \frac{\partial p_0}{\partial z} = -g ,
\end{equation}
it follows that the total differentials for $\Pi_m$ and ${\cal B}_R$ at fixed $t$ are particularly simple,
\begin{equation}
    {\rm d}\Pi_m = (T_{\star} - T_R) {\rm d}\eta 
    + \nu_{\star} \frac{\partial p_m}{\partial r_{\star}} {\rm d}r_{\star} ,
\end{equation}
\begin{equation}
    {\rm d}{\cal B}_R = T_R \, {\rm d}\eta ,
\end{equation}
where $T_R = T(\eta,p_R)$, $T_{\star} = T(\eta,p_{\star})$ and 
$\nu_{\star} =\nu(\eta,p_{\star})=\nu_m (r_{\star},z_{\star},t)$. These relations will prove useful to construct a full energy cycle in the next section.  
}

The mean kinetic energy of the reference vortex is simply equal to $v_{\star}^2/2$. In contrast to the case $f=0$ considered by \citet{Andrews2006}, the case of a finite rotation rate $f\ne 0$ considered in this paper gives rise to a background radial distribution of angular momentum $M_R(r) = f r^2/2$ that introduces a radial restoring inertial force, in the same way that a statically stable vertical gradient of entropy introduces a vertical restoring buoyancy force. As mentioned previously, a reference equilibrium radius $r_R$ (for which $v_R = 0$) can then be defined as the solution of $M = f r_R^2/2$. As a result, we may write $v_{\star}$ as
\begin{equation}
      v_{\star} = \frac{M}{r_{\star}} - \frac{fr_{\star}}{2} = \frac{f r_R^2}{2 r_{\star}} - \frac{f r_{\star}}{2} = 
      \frac{f(r_R + r_{\star})(r_R-r_{\star})}{2 r_{\star}},
      \label{vstar_definition} 
\end{equation}
which in turn implies for the kinetic energy $v_{\star}^2/2$ 
\begin{equation}
      \frac{v_{\star}^2}{2} = \frac{f^2 (r_R+r_{\star})^2 (r_R - r_{\star})^2}{8 r_{\star}^2} .
      \label{vstar_as_work}
\end{equation}
Eq. (\ref{vstar_as_work}) shows that $v_{\star}^2/2$ is quadratic in the displacement amplitude $\delta r = r_{\star} - r_R$ from the equilibrium position $r_R$ even for finite-amplitude $\delta r$. Moreover, Eq. (\ref{vstar_definition}) shows that creating a cyclonic circulation ($v_{\star}>0$ if $f>0$) requires $r_R>r_{\star}$, and hence the compression of the equilibrium constant angular momentum surfaces.  In order to express $v_{\star}^2/2$ as the work against the inertial restoring force, we may use (\ref{v2intermsofM}) to write it in the form:
\begin{equation}
    \frac{v_{\star}^2}{2} = \frac{v_{\star}^2}{2} - \frac{v_R^2}{2} =  \mu (\chi_{\star} - \chi_R) + \frac{f^2}{16} \left ( \frac{1}{\chi_{\star}} - \frac{1}{\chi_R} \right ) .
\end{equation} 
making use of the fact that $v_R^2 = \mu \chi_R + f^2/(16\chi_R) - f\sqrt{\mu}/2 = 0$, which in turn may be written as the following integrals:
\begin{equation}
     \frac{v_{\star}^2}{2} 
     = \int_{r_R}^{r_{\star}} \left ( \mu - \mu_R(r') \right ) \,
     \frac{\partial \chi}{\partial r}(r')\,{\rm d}r' = 
     \int_{\chi_R}^{\chi_{\star}} 
     \left ( \mu - \mu_R(\chi') \right ) 
     \,{\rm d}\chi' .
     \label{vstar_as_integral} 
\end{equation}
We propose to call the right-hand side of (\ref{vstar_as_work}) and its integral expression (\ref{vstar_as_integral}) the centrifugal potential energy. Eq. (\ref{vstar_as_integral}) makes it clear that the restoring inertial force experienced by the fluid parcel as it experiences a radial displacement is $-(\mu-\mu_r)\partial \chi/\partial r = (\mu-\mu_R)/r^3$, consistent with \citet{Emanuel1994}. For small displacements $\delta r$, so that $r_R \approx r_{\star}$, the above expression approximates to $v_{\star}^2/2 \approx f^2(r_{\star}-r_R)^2/2$, which suggests that the natural frequency of the radial waves is $f$, as would be the case for standard inertial waves. These results clearly establish that $\Pi_k$ and $\Pi_e$ are the natural generalisations of $v_{\star}^2/2$ and $\Pi_m$ for constructing the actual state from a non-resting reference state instead of a resting one.

\section{Energetics and dynamics of vortex growth due to diabatic effects}

\textcolor{violet}{The results of the previous section establish that for a symmetrically stable reference vortex, the total energy of the unbalanced part ${\bf u}_s^2/2 + \Pi_1 + \Pi_k + \Pi_e$, the total energy of the balanced part $v_{\star}^2/2 + \Pi_m$ and the background potential energy ${\cal B}_R$ are all individually conserved for purely adiabatic and inviscid axisymmetric disturbances, owing to the latter two forms of energy being functions of $\eta$ and $M$ only and hence material invariants.} In that case, ${\bf b}_e$ acts a restoring force giving rise to a complex superposition of internal and inertial/centrifugal waves, as discussed by \citet{Emanuel1994} for instance. As long as the conditions for symmetric stability (\ref{stability_pie}) and (\ref{pik_positive}) \textcolor{violet}{hold}, any \textcolor{violet}{reversible} transfer between the eddy and mean energies is forbidden. \textcolor{violet}{In this section, we seek to understand how such constraints might be relaxed as well as the nature of the additional energy conversions arising from the presence of sinks and sources of $M$ and $\eta$. A motivating issue related to the problem of TC intensification is how a positive diabatic source of heating $D\eta/Dt = \dot{q}/T > 0$ and negative sink of angular momentum $DM/Dt<0$ combine to intensify an incipient warm core cyclonic seed vortex, keeping in mind the intrinsic limitations of a dry framework.} The discussion of non-axisymmetric effects, which is significantly more involved, is left to a future study.

\subsection{Standard energetics viewpoint}

\textcolor{violet}{Prior to delving into more sophisticated considerations of available energetics, it is useful to review what the standard energetics viewpoint can tell us about vortex intensification, e.g., \citet{Smith2018}.} To that end, we place ourselves in a Northern Hemisphere-like situation $(v>0,f>0)$. \textcolor{violet}{A common approach is to} consider separate evolution equations for the azimuthal kinetic energy $v^2/2$ and the rest of the flow as follows:
\begin{equation}
    \rho \frac{D}{Dt} \left ( \frac{{\bf u}_s^2}{2} + \Phi + h - \frac{p}{\rho} \right )  + \nabla \cdot (p {\bf u}_s) = \rho {\bf u}_s \cdot {\bf D}_s + \rho \dot{q} + \left ( f + \frac{v}{r} \right ) \rho u v , 
    \label{us_energetics}
\end{equation}
\begin{equation}
    \rho \frac{D}{Dt} \frac{v^2}{2} = - \left ( f + \frac{v}{r} \right ) \rho u v + \rho v D_v .
    \label{v_energetics}
\end{equation}
 Since the dissipation term $v D_v$ presumably acts as a brake on $v$, Eq. (\ref{v_energetics}) demonstrates that because $v>0$ by design, the radial velocity must be \textcolor{violet}{dominantly} negative $(u<0)$ in order for \textcolor{violet}{the overall kinetic energy of the azimuthal circulation $v^2/2$ to increase, as this is required for the conversion term $-(f+v/r)uv$ to be positive. This condition is of course well known and observed in the cyclonic circulations of numerically simulated TCs. According to Eqs. (\ref{us_energetics}-\ref{v_energetics}), there are two main energy reservoirs that the azimuthal kinetic energy $v^2/2$ can potentially tap into: 1) the kinetic energy of the secondary circulation ${\bf u}_s^2/2$ and 2) the part of the potential energy $\Phi + h-p/\rho$ available for conversions into kinetic energy, the so-called APE. Ultimately, however, the main overall source of energy for a TC are the enthalpy fluxes exchanged at the air-sea interface. As these can only create APE, there are really only two plausible energy pathways for vortex intensification, schematically:
\begin{equation}
   {\rm Heating} \quad \rightarrow \quad {\rm APE}
   \quad \rightarrow \quad \frac{{\bf u}_s^2}{2} \quad
   \quad \rightarrow  \quad \frac{v^2}{2}, 
   \label{energy_pathways_1} 
\end{equation}
\begin{equation} 
    {\rm Heating} \quad \rightarrow \quad  {\rm APE} \quad 
    \rightarrow \quad \frac{v^2}{2} ,
    \label{energy_pathway_2} 
\end{equation} 
which states that the APE is either converted directly into $v^2/2$ or indirectly via preliminary creation of the secondary circulation kinetic energy ${\bf u}_s^2/2$. 
Theoretically, the problem of understanding TC intensification therefore amounts, at least partly, to elucidating the physical mechanisms giving rise to the secondary circulation, as well as the relative importance of the ${\bf u}_s^2/2$ and available potential energy reservoirs as the sources of energy for $v^2/2$. So far, the main theoretical frameworks for discussing the physics of the secondary circulation have been the Sawyer-Eliassen equations forming the basis of purely balanced models of TC evolution, and the concept of generalised buoyancy force, as reviewed in \citet{Montgomery2017}. } 

\textcolor{violet}{Because the temporal behaviour of the azimuthal kinetic energy $v^2/2$ may not always be representative of that of the maximum wind $v_{max}$, the metric that is more commonly used to study TC intensification, it is also useful to seek insights into the intensification of $v_{max}$ from} the angular momentum conservation equation (\ref{angular_momentum}) --- which has essentially the same information content as (\ref{v_energetics}) --- written in 
Eulerian form:
\begin{equation}
    \frac{\partial M}{\partial t} = - u \frac{\partial M}{\partial r} - w \frac{\partial M}{\partial z} + r D_v .
    \label{momentum_condition} 
\end{equation}
If the distribution of $M$ is such that $\partial M/\partial r>0$ and $\partial M/\partial z<0$, as is seen to be the case for the analytical reference vortex case described in Appendix \ref{analytical_solution} and illustrated in Fig. \ref{fig:lifting_illustration}, Eq. \ref{momentum_condition} makes it clear that both $u<0$ and $w>0$ will contribute to the local intensification of $M$ and hence of $v$. The understanding of axisymmetric TC intensification therefore boils down to understanding how viscous and diabatic effects cooperate to drive an upward and radially inward secondary circulation at low levels near the eyewall. 

\textcolor{violet}{In the following, the azimuthal circulation is regarded as the sum of a balanced and unbalanced parts $v = v_{\star} + v''$ (Lagrangian viewpoint) or $v = v_m + v'$ (Eulerian viewpoint), as is commonly done in the TC literature. Evidence from the literature suggests that both the balanced and unbalanced parts are important to explain observed TC intensification, the unbalanced part being especially important in the boundary layer, e.g., \citet{Bui2009}. The unbalanced part of a vortex has also been shown to be important to account for so-called superintensities, e.g., see \citet{Rousseau-Rizzi2019} for a recent review of the issue and references. Now, because $v' = v-v_m = (M-M_m)/r = (\mu-\mu_m)/(r(M+M_m))$, any increase in $v'$ must result from the creation of a positive anomaly $\mu' = \mu-\mu_m>0$ and hence from an increase in the mechanical energy reservoir $\Pi_k$, the only one that increases when $|\mu-\mu_m|$ increases, as further discussed below. 
}

\textcolor{violet}{
\subsection{Separation of the circulation into diabatic and adiabatic components}
%\subsubsection{General approach to defining diabatic and adiabatic circulations} 
In the TC literature, the secondary (or transverse) circulation is traditionally regarded as purely diabatic and diagnostic, as this is how it enters the Sawyer-Eliassen equations forming the basis of purely balanced models of TC evolution. In reality, the secondary circulation also possesses an adiabatic and prognostic component. It is therefore essential in our framework to explain how to separate the diabatic and diagnostic component of the secondary circulation from its adiabatic and prognostic component.}

\textcolor{violet}{To proceed, let us first remark that in all regions where the axisymmetric potential vorticity $Q = J/(\rho r)$ does not change sign --- where $J = \partial (M,\eta)/\partial (r,z)$ --- a one-to-one (time-dependent) map between physical space $(r,z)$ and the space of material invariants $(M, \eta)$ can be constructed from the expressions for the specific angular momentum $M = M(r,z,t)$ and specific entropy $\eta = \eta(r,z,t)$. Indeed, inverting such relations allows one to regard $r$ and $z$ as functions of $M$ and $\eta$,
\begin{equation}
       r = \tilde{r}(M,\eta,t), \qquad z = \tilde{z}(M,\eta,t) .
       \label{inverse_relations} 
\end{equation}
From taking the total material derivative of (\ref{inverse_relations}), the components of the secondary circulation may be written in the following form
\begin{equation}
       u = \frac{Dr}{Dt} = \underbrace{\frac{\partial \tilde{r}}{\partial M} \dot{M}
       + \frac{\partial \tilde{r}}{\partial \eta} \dot{\eta}}_{u_d} 
       + \underbrace{\frac{\partial \tilde{r}}{\partial t}}_{u_{ad}}, \qquad
       w = \frac{Dz}{Dt} = \underbrace{\frac{\partial \tilde{z}}{\partial M} \dot{M}
       + \frac{\partial \tilde{z}}{\partial \eta} \dot{\eta}}_{w_d} +
       \underbrace{\frac{\partial \tilde{z}}{\partial t}}_{w_{ad}} ,
       \label{secondary_circulation_decomposition} 
\end{equation}
where it is easily established that 
\begin{equation}
     \frac{\partial \tilde{r}}{\partial M} = 
     \frac{1}{J} \frac{\partial \eta}{\partial z}, \qquad
     \frac{\partial \tilde{r}}{\partial \eta} = 
     - \frac{1}{J} \frac{\partial M}{\partial z} , \qquad
     \frac{\partial \tilde{z}}{\partial M} = 
     - \frac{1}{J} \frac{\partial \eta}{\partial r}, \qquad
     \frac{\partial \tilde{z}}{\partial \eta} = 
     \frac{1}{J} \frac{\partial M}{\partial r}  .
     \label{derivatives_expressions} 
\end{equation}
Eqs. (\ref{secondary_circulation_decomposition}) establish that the transverse circulation $(u,w)$ can naturally be partitioned as the sum of a purely diagnostic `diabatic' component $(u_d,w_d)$ that depends only on the instantaneous values of $\dot{M}$ and $\dot{\eta}$ plus a prognostic `adiabatic' component
\begin{equation}
       (u_{ad},w_{ad})  = \left ( \frac{\partial \tilde{r}}{\partial t}, 
       \frac{\partial \tilde{z}}{\partial t} \right ) .
\end{equation}
By making use of (\ref{derivatives_expressions}) in the budget equations $D\eta/Dt = \dot{\eta}$ and $DM/Dt = \dot{M}$, it is easily established that the `diabatic' and `adiabatic' components of $(u,w)$ satisfy
\begin{equation}
     u_d \frac{\partial F}{\partial r} + w_d \frac{\partial F}{\partial z} = \dot{F}, \qquad \frac{D_{ad} F}{Dt} = 0 , 
\end{equation}
for $F = (\eta,Q)$, where $D_{ad}/Dt = \partial/\partial t + u_{ad} \partial/\partial r + w_{ad} \partial/\partial z$ is the Lagrangian derivative defined by the adiabatic component of the circulation. In other words, $\eta$ and $M$ are materially conserved along the adiabatic trajectories defined by $(u_{ad},w_{ad})$. 
Note that the above considerations represent no more than a general extension of the well-known isentropic analysis, which exploits the fact that in a statically stable stratified fluid, entropy increases with height, thus allowing it to be used as a generalised vertical coordinate. Except in the boundary layer, the adiabatic component $(u_a,w_a)$ of the secondary circulation is expected to dominate over its diabatic component $(u_d,w_d)$. It is important to realise that the terminology `adiabatic' does not imply that $(u_a,w_a)$ is not affected by friction or sources/sinks of entropy, only that it is not instantaneously or diagnostically determined by the latter. The procedure leading to the Sawyer-Eliassen equations presumably causes the finite adjustment time scales by which $(u_a,w_a)$ responds to the sinks/sources of angular momentum and entropy to become infinitely fast, thus transforming $(u_a,w_a)$ into a diagnostic quantity. Indeed, it is physically expected that the Sawyer-Eliassen equations should be a diagnostic description of the full transverse circulation, not just of $(u_d,w_d)$, since the latter is often negligible as compared to $(u_a,w_a)$. 
}

\textcolor{violet}{
\subsection{Diabatic nature of the secondary reference vortex circulation} 
Although the above partitioning into a diabatic and adiabatic component is exact and rigorous, it is not easily exploited. In the following, we consider an alternative approach that is easier to use in practice. Because the balanced reference vortex state is by construction assumed to be obtainable from the actual state by means of an adiabatic and inviscid re-arrangement of mass, it is possible to show that the part of the secondary circulation associated with the Lagrangian motion of a parcel's reference position $(r_{\star},z_{\star})$,  
\begin{equation}
       u_{\star} = \frac{Dr_{\star}}{Dt}, \qquad
       w_{\star} = \frac{Dz_{\star}}{Dt} ,
\end{equation}
is also a purely diagnostic function of the sinks/sources of $M$ and $\eta$, similarly to the diabatic component $(u_d,w_d)$ defined above. To show this, let us take the total material derivative of the mathematical relations (\ref{reference_state_system}) defining $(r_{\star},z_{\star})$, repeated here with time dependence retained, 
\begin{equation}
 M_m(r_{\star},z_{\star},t) = M, \qquad \eta_m(r_{\star},z_{\star},t) = \eta ,
 \label{generalised_LNB_equations} 
\end{equation}
which leads to the following system of equations: 
\begin{equation}
     u_{\star} \frac{\partial \eta_m}{\partial r} 
     + w_{\star} \frac{\partial \eta_m}{\partial z} = \frac{D\eta}{Dt} 
     - \frac{\partial \eta_m}{\partial t} 
= u_d \frac{\partial \eta}{\partial r} 
+ w_d \frac{\partial \eta}{\partial z} - \frac{\partial \eta_m}{\partial t}, 
     \label{ustar_system}
\end{equation}
\begin{equation}
     u_{\star} \frac{\partial M_m}{\partial r} + w_{\star} \frac{\partial M_m}{\partial z} = \frac{DM}{Dt} - \frac{\partial M_m}{\partial t} 
     = u_d \frac{\partial M}{\partial r} + w_d 
     \frac{\partial M}{\partial z} - \frac{\partial M_m}{\partial t} ,
     \label{wstar_system}
\end{equation}
in which all the $m$-suffixed quantities are evaluated at $(r_{\star},z_{\star},t)$. Provided that the Jacobian $J_0 = \partial (M_m,\eta_m)/\partial (r,z)$ differs from zero, such a system can be inverted for $u_{\star}$ and $w_{\star}$ as follows:
\begin{equation}
    u_{\star}  = \frac{1}{J_0} \left \{ \frac{\partial \eta_m}{\partial z}
    \left ( \frac{DM}{Dt} - \frac{\partial M_m}{\partial t} \right )  
    - \frac{\partial M_m}{\partial z} \left ( \frac{D\eta}{Dt}
    - \frac{\partial \eta_m}{\partial t} \right ) \right \} ,
    \label{ustar_solution}
\end{equation}
\begin{equation}
    w_{\star} = \frac{1}{J_0} \left \{ - \frac{\partial \eta_m}{\partial r}
 \left ( \frac{DM}{Dt} - \frac{\partial M_m}{\partial t} \right ) +
    \frac{\partial M_m}{\partial r} \left (
    \frac{D\eta}{Dt} - \frac{\partial \eta_m}{\partial t} \right )  \right \} . 
    \label{wstar_solution} 
\end{equation}
The circulation $(u_{\star},w_{\star})$ is closely related to the diabatic circulation $(u_d,w_d)$ defined above, 
$$
     u_{\star} =  \frac{1}{J_0} 
     \left [ \left ( \frac{\partial \eta_m}{\partial z}
     \frac{\partial M}{\partial r} - \frac{\partial M_m}{\partial z}
     \frac{\partial \eta}{\partial r} \right ) u_d 
     + \left ( \frac{\partial \eta_m}{\partial z}
     \frac{\partial M}{\partial z} - \frac{\partial M_m}{\partial z}
     \frac{\partial \eta}{\partial z} \right ) w_d \right . 
$$
\begin{equation} 
\left . 
     + \frac{\partial M_m}{\partial z} \frac{\partial \eta_m}{\partial t}
      - \frac{\partial \eta_m}{\partial z} \frac{\partial M_m}{\partial t}
      \right ] 
\end{equation}
$$
     w_{\star}  = \frac{1}{J_0} 
     \left [ \left ( \frac{\partial M_m}{\partial r} 
     \frac{\partial \eta}{\partial r} - \frac{\partial \eta_m}{\partial r}
      \frac{\partial M}{\partial r} \right ) u_d 
      + \left ( \frac{\partial M_m}{\partial r} 
      \frac{\partial \eta}{\partial z} - \frac{\partial M}{\partial z}
      \frac{\partial \eta_m}{\partial r} \right ) w_d  \right . 
$$
\begin{equation}
 \left . 
    \frac{\partial \eta_m}{\partial r} \frac{\partial M_m}{\partial t}
    - \frac{\partial M_m}{\partial r} \frac{\partial \eta_m}{\partial t}
         \right ] . 
\end{equation}
This expression shows that $(u_{\star},w_{\star})$ reduces to $(u_d,w_d)$ in the case where $M \approx M_m$, $\eta \approx \eta_m$ and the time derivative of $M_m$ and $\eta_m$ can be neglected. Because $\eta_m$ and $M_m$ depend on time only if diabatic/viscous effects are retained, it follows that their partial derivatives $\partial \eta_m/\partial t$ and $\partial M_m/\partial t$ must be solely functions of the diabatic sinks and sources of entropy and angular momentum $D\eta/Dt = \dot{\eta}$ and $DM/Dt = \dot{M}$. This can be explicitly established in the case of the Lorenz reference state (see \citet{Novak2018} p. 1895 for instance). This result is important, because it establishes that $u_{\star}$ and $w_{\star}$ represent purely diagnostic quantities, which in principle can be evaluated at any time $t$ from the known distribution of $\dot{\eta}$ and $\dot{M}$.}

\textcolor{violet}{The diabatic secondary circulation $(u_{\star},w_{\star})$ defined above shares several important similarities and differences with the secondary circulation $(u_{se},w_{se})$ entering the Sawyer-Eliassen equations that form the basis for purely balanced models of TC evolution (see \citet{Montgomery2017} for a review). As regards to similarities, both circulations share the property of being purely diagnostic functions of $\dot{\eta}$ and $\dot{M}$. Moreover, the problems determining both circulations are mathematically well-posed only if the conditions for symmetric and inertial stability are satisfied. In the case of $(u_{\star},w_{\star})$, these conditions reduce to the requirement that the Jacobian 
\begin{equation}
      J_0 = \frac{\partial (M_m,\eta_m)}{\partial (r,z)} = 
      r \rho_m Q_m ,
\end{equation}
(which is proportional to the potential vorticity of the balanced reference state) remains single-signed throughout the domain, as this is the condition needed for (\ref{ustar_system}-\ref{wstar_system}) to be invertible. Such a condition is also sufficient in general to guarantee that the Sawyer-Eliassen equations are elliptic and invertible. As regards to differences, a key one is that unlike the Sawyer-Eliassen equations, the system  (\ref{ustar_system}-\ref{wstar_system}) determining $(u_{\star},w_{\star})$ is not elliptic and therefore much simpler to solve, since it can be trivially inverted to express $u_{\star}$ and $w_{\star}$ in terms of local and non-local functions of $\dot{\eta}$ and $\dot{M}$, as shown by (\ref{ustar_solution}-\ref{wstar_solution}). Perhaps the main key difference, however is due to the fact that by construction, the physics of $(u_{se},w_{se})$ attempts to capture elements pertaining to both balanced and unbalanced evolution, whereas that of $(u_{\star},w_{\star})$ pertains solely to the balanced evolution. Physically, one part of $(u_{se},w_{se})$ that a priori pertains to unbalanced evolution is that controlled by the rate of generation of the generalised buoyancy force, as first shown by \citet{Smith2005}. This is because --- as confirmed by the present results --- the generalised buoyancy force only pertains to the evolution of the unbalanced part of the vortex (as well as to the `adiabatic' part $(u_a,w_a) = (u-u_{\star},w-w_{\star})$ of the secondary circulation); it is irrelevant to the physics of $(u_{\star},w_{\star})$. \citet{Smith2005}'s results are therefore important theoretically, because they make it clear that purely balanced models of TC evolution are actually capable of capturing the physics of both the balanced and unbalanced parts of a vortex (whether such models also respect the relative importance of each part is another matter beyond the scope of this paper). Because the balanced and unbalanced parts of a vortex are affected differently by the sinks/sources of $\eta$ and $M$, it follows that the ways in which $(u_{\star},w_{\star})$ and $(u_{se},w_{se})$ depend on $\dot{\eta}$ and $\dot{M}$ must also be different, although a full discussion of such differences is beyond the scope of this paper. Finally, it seems useful to point out that experience with balanced models of TC evolution based on the use of the Sawyer-Eliassen equations has shown that balanced TCs often become inertially and/or symmetrically unstable resulting in the Sawyer-Eliassen equations losing their elliptic character and invertibility, e.g., \citet{Smithetal2018}, in which case some form of regularization may be used to carry on the integration \cite{Wang2019}. How to solve for $(u_{\star},w_{\star})$ when $J_0=0$ is an interesting and important issue but beyond the scope of this paper. 
} 

\textcolor{violet}{
\subsection{Evolution and energetics of the diabatic azimuthal balanced circulation}
We first seek insights into the evolution of the balanced part of the azimuthal vortex circulation $v_m(r,z,t)$, which in the TC literature is generally considered to dominate TC intensification. As mentioned previously, the balanced reference state is assumed to be obtainable from the actual state by means of an invisicid and adiabatic re-arrangement of mass. Even though how to construct such a state in practice is not understood, we show in the following that this does preclude the derivation of explicit theoretical results about how the balanced part of the vortex is affected by sinks and sources of $M$ and $\eta$.  To proceed, it is sufficient to focus on $v_m(r,z,t)$ evaluated at a parcel's reference position, viz., 
\begin{equation}
       v_{\star} = v_m (r_{\star},z_{\star},t) = 
       \frac{M}{r_{\star}} - \frac{fr_{\star}}{2}, 
       \label{vstar_expression}
\end{equation}
owing to the one-to-one map that is assumed to exist between the actual state and reference state. To make progress, let us take the substantial derivative of  (\ref{vstar_expression}). The result can be written either in Lagrangian form as
\begin{equation}
     \frac{Dv_{\star}}{Dt} = \frac{1}{r_{\star}}  \frac{DM}{Dt}
      - u_{\star} \left ( \frac{M}{r_{\star}^2} + \frac{f}{2}  \right ),
     \label{prognostic_vstar} 
\end{equation}
or alternatively in Eulerian form as
$$
    \frac{\partial v_{\star}}{\partial t} 
    = - u_{\star} \frac{\partial v_{\star}}{\partial r_{\star}} 
      - w^{\star} \frac{\partial v_{\star}}{\partial z_{\star}} +
      \frac{1}{r_{\star}}  \frac{DM}{Dt}
      - u_{\star} \left ( \frac{M}{r_{\star}^2} + \frac{f}{2}  \right ) 
$$
\begin{equation}
     = -  u_{\star} \left [ f + \frac{1}{r_{\star}} 
     \frac{\partial (r_{\star} v_{\star})}{\partial r_{\star}} \right ]   
     - w_{\star} \frac{\partial v_{\star}}{\partial z_{\star}} + 
     \frac{1}{r_{\star}} \frac{DM}{Dt} ,
     \label{eulerian_vstar} 
\end{equation}
where the last part of the equation follows from using (\ref{vstar_expression}) to rewrite the term $M/r_{\star}^2 + f/2$ in terms of $v_{\star}$. The term within square brackets in (\ref{eulerian_vstar}) can be recognised as the vertical component of the reference absolute vorticity, and is presumably dominantly positive in TCs, while the term $\partial v_{\star}/\partial z <0$ is expected to be negative for a warm core cyclonic vortex, since maximum winds are near the ground. Since $DM/Dt<0$ by assumption, it follows that for $v_{\star}$ to locally intensify generally, the conditions $u_{\star} < 0$ and $w_{\star} > 0$ are generally required, which are the same that need to be satisfied by the full secondary circulation ${\bf u}_s = {\bf u}_{\star} + {\bf u}_a$.}

\textcolor{violet}{
As seen above, the way ${\bf u}_{\star}$ is controlled by sinks and sources of $\eta$ and $M$ is predicted by Eqs. (\ref{ustar_solution}-\ref{wstar_solution}). To get a sense of what these relations predict, let us consider the case where the non-local temporal variations of $\eta_m$ and $M_m$ can be neglected relative to the local values $\dot{\eta}$ and $\dot{M}$, which is the only case that is analytically tractable. Visual inspection of Fig. \ref{fig:lifting_illustration} shows that the signs of the partial derivatives of $M_m$ and $\eta_m$ satisfy 
\begin{equation}
     \frac{\partial \eta_m}{\partial z} > 0, \qquad 
     \frac{\partial M_m}{\partial z} < 0, \qquad
     \frac{\partial \eta_m}{\partial r}< 0, \qquad
     \frac{\partial M_m}{\partial r} > 0 . 
\end{equation}
As a result, (\ref{ustar_solution}-\ref{wstar_solution}) imply
\begin{equation}
    u_{\star}  \approx \underbrace{
    \frac{1}{J_0} \frac{\partial \eta_m}{\partial z}
    \frac{DM}{Dt}}_{<0} \qquad 
    \underbrace{- \frac{1}{J_0} \frac{\partial M_m}{\partial z} 
    \frac{D\eta}{Dt}}_{>0} ,
    \label{approximate_ustar_solution}
\end{equation}
\begin{equation}
    w_{\star} \approx \underbrace{
    - \frac{1}{J_0}\frac{\partial \eta_m}{\partial r}
 \frac{DM}{Dt}}_{<0}  \qquad +
    \underbrace{\frac{1}{J_0} \frac{\partial M_m}{\partial r} 
    \frac{D\eta}{Dt}}_{>0} . 
    \label{approximate_wstar_solution} 
\end{equation}
Eqs. (\ref{approximate_ustar_solution}-\ref{approximate_wstar_solution}) show that a positive source of diabatic heating $D\eta/Dt<0$ and sink of angular momentum $DM/Dt<0$ oppose each other to achieve the conditions $u_{\star}>0, w_{\star}>0$. Indeed, Eq. (\ref{approximate_ustar_solution}) shows that friction must dominate over diabatic heating in order to achieve inward $u_{\star}$, whereas Eq. (\ref{approximate_wstar_solution}) shows that diabatic heating must dominate over friction to achieve upward $w_{\star}>0$. The fact that friction appears to be essential for the intensification of the balanced reference vortex is reminiscent of \citet{Charney1964}'s idea that friction should be regarded as playing a dual role in promoting both the intensification and weakening of TCs, which is central to the so-called CISK paradigm. These results are important, because we show in the following subsection that the way that friction and diabatic heating tend to control the adiabatic component ${\bf u}_a$ is very different from the way that they control ${\bf u}_{\star}$. For instance, we show that diabatic heating and friction tend to promote a radially inward and outward $u_a$ respectively, which is the reverse of that for $u_{\star}$. On a last note, it is also interesting to remark that in Eq. (\ref{wstar_system}), the term $\partial M_m/\partial r = r^3 I^2/(2 M_m)$ is proportional to the inertial stability parameter $I^2$, which is consistent with the idea that the efficiency of diabatic heating in driving TC intensification increases with $I^2$, as discussed in \citet{Schubert1982} for instance. 
}

\textcolor{violet}{To conclude this subsection, let us seek to clarify the energetics of $v_{\star}$. To that end, let us multiply both sides of Eq. (\ref{prognostic_vstar}) by $v_{\star}$, accounting for (\ref{vstar_expression}) and the equations defining the balanced reference vortex, thus leading to
\begin{equation}
     \frac{D}{Dt} \frac{v_{\star}^2}{2} 
     = \frac{v_{\star}}{r_{\star}} \frac{DM}{Dt}
       - u_{\star} \left ( \frac{M^2}{r_{\star}^3} - \frac{f^2 r_{\star}}{4} \right ) = \frac{v_{\star}}{r_{\star}} 
       \frac{DM}{Dt} - \frac{u_{\star}}{\rho_{\star}} 
       \frac{\partial p_{\star}}{\partial r_{\star}} .
       \label{kinetic_vstar} 
\end{equation}
 To clarify the nature of the energy conversions entering (\ref{kinetic_vstar}), let us derive the evolution equation for $\Pi_m$ and ${\cal B}_R$, obtained by taking the substantial derivatives of (\ref{mean_APE_density}) and (\ref{background_potential_energy}) respectively: 
\begin{equation}
      \frac{D\Pi_m}{Dt} = 
      \left ( \frac{T_{\star} - T_R}{T} \right ) \dot{q}
      + \frac{u_{\star}}{\rho_{\star}} \frac{\partial p_{\star}}
      {\partial r_{\star}}
      + \frac{1}{\rho_{\star}} \frac{\partial p_{\star}}{\partial t}
      - \frac{1}{\rho_R} \frac{\partial p_R}{\partial t} , 
      \label{pim_evolution} 
\end{equation}
\begin{equation}
     \frac{D{\cal B}_R}{Dt} = \frac{T_R}{T} \dot{q} 
     + \frac{1}{\rho_R} \frac{\partial p_R}{\partial t}.  
     \label{br_evolution} 
\end{equation}
Eq. (\ref{pim_evolution}) makes it clear that the term $u_{\star}/\rho_{\star} \partial p_{\star}/\partial r_{\star}$ represents an energy conversion between $\Pi_m$ and $v_{\star}^2/2$, $\Pi_m$ itself being created by diabatic heating at the rate $(T_{\star}-T_R)\dot{q}/T$. Eq. (\ref{br_evolution}) also shows that some energy exchange between $\Pi_m$ and ${\cal B}_R$ is possible via the term $\rho_R^{-1} \partial p_R/\partial t$, while next subsection will make it clear that the term $\rho_{\star} \partial p_{\star}/\partial t$ allows for some energy exchange between the mean APE $\Pi_m$ and the unbalanced eddy APE $\Pi_e$. Eqs. (\ref{pim_evolution}) therefore states that diabatic heating (which for TC originates from surface enthalpy fluxes) represents the ultimate source of energy for the kinetic energy of the balance reference state $v_{\star}/2$ via the creation of mean APE $\Pi_m$, which is consistent with the hypothesised energy pathway \ref{energy_pathway_2}. However, in order for $\Pi_m$ to be convertible into $v_{\star}^2/2$, an inward radial component $u_{\star}<0$ is essential, which is possible only if friction dominates over diabatic heating in (\ref{approximate_ustar_solution}). The resulting energy cycle is illustrated in the bottom part of Fig. \ref{fig:energy_cycle}. 
}

\textcolor{violet}{
\subsection{Driving and energetics of the unbalanced part of the vortex}
We now turn to the problem of clarifying the physics and energetics of the unbalanced part of the vortex intensification. To that end, we need to understand the circumstances causing the adiabatic part of the secondary circulation ${\bf u}_a = {\bf u}_s - {\bf u}_{\star}$ to be radially inward and vertically upward. Unlike ${\bf u}_{\star}$, which is a purely diagnostic function of the sinks/sources of $\eta$ and $M$, ${\bf u}_a$ is a prognostic quantity whose behaviour is therefore determined by the forces acting on it. From the results of the previous section, we expect those forces to be dominated by the generalised inertial/buoyancy force ${\bf b}_e$. To proceed, let us construct the prognostic equation for ${\bf u}_s$ by subtracting  (\ref{gradient_wind_balance}) from (\ref{radial}-\ref{vertical}), which yields} 
\begin{equation}
     \frac{D{\bf u}_s}{Dt} = {\bf b}_e  -\frac{1}{\rho}\nabla p' - \nu' \nabla p_m + {\bf D}_s .
     \label{secondary_circulation} 
\end{equation}
\textcolor{violet}{
Since ${\bf u}_{\star}$ can be regarded as known, it is useful to rewrite (\ref{secondary_circulation}) with the parts of $D{\bf u}_s/Dt$ involving ${\bf u}_{\star}$ on the right-hand side so as it make it appears as a prognostic equation for ${\bf u}_a$,  
\begin{equation}
     \frac{D_a {\bf u}_a}{Dt} = {\bf b}_e - \frac{1}{\rho} \nabla p' 
      - \nu' \nabla p_m + {\bf D}_s - \frac{D{\bf u}_{\star}}{Dt} 
      - ({\bf u}_{\star} \cdot \nabla ) {\bf u}_a ,
      \label{secondary_circulation_bis} 
\end{equation}
where $D_a/Dt = \partial_t + {\bf u}_a \cdot \nabla$ is the Lagrangian derivative following ${\bf u}_a$.} From (\ref{generalised_buoyancy}), the radial and vertical components of ${\bf b}_e$ may be written explicitly as follows:  
\begin{equation}
    b_e^{(r)} = -(\nu_h - \nu_m ) \frac{\partial p_m}{\partial r} 
    + \frac{(\mu-\mu_m)}{r^3} = 
    -\rho_m (\nu_h -\nu_m) \left ( f + \frac{v_m}{r} \right ) v_m 
    + \frac{(\mu-\mu_m)}{r^3} , 
    \label{radial_be}
\end{equation}
\begin{equation}
     b_e^{(z)} = -(\nu_h - \nu_m) \frac{\partial p_m}{\partial z} = \rho_m g (\nu_h -\nu_m) .
     \label{vertical_be}
\end{equation}
\textcolor{violet}{Because of the complexity of the problem, it is difficult in general to make definite theoretical statements about the magnitude and sign of the various terms in the right-hand side of (\ref{secondary_circulation_bis}) with the exception of ${\bf b}_e$. Because for an intensifying vortex, $(\mu-\mu_m)/r^3>0$ acts to drive $u_a$ outward, 
}
Eq. (\ref{radial_be}) shows that a necessary condition for the radial component of ${\bf b}_e$ to point radially inward is that the fluid parcels be positively buoyant 
\textcolor{violet}{relative to the reference vortex state},
\begin{equation}
        \nu_h - \nu_m > 0,
        \label{buoyant_condition} 
\end{equation}
in which case (\ref{vertical_be}) shows that $b_e^{(z)}$ will also point upward, \textcolor{violet}{as previously discussed by \citet{Smith2005}}. By definition, $\nu_h = \nu(\eta,p_m(r,z))$ and $\nu_m = \nu(\eta_m(r,z),p_m(r,z))$ so that 
\begin{equation}
       \nu_h - \nu_m \approx \Gamma (\eta(r,z,t) - \eta_m(r,z)) 
\end{equation}
is proportional to the local entropy anomaly $\eta' = \eta - \eta_m$ (we have neglected the time variation of the reference variables, but these can be retained if desired). Since $\Gamma = \partial \nu/\partial \eta(\eta_m,p_m) >0$ in general, the creation of a positive specific volume anomaly requires a sustained diabatic source of entropy to increase $\eta$. 
\begin{comment}
As discussed by \citet{Smith2005}, whether (\ref{buoyant_condition}) is satisfied depends critically on the choice of reference state used to define buoyancy. For instance, \citet{Braun2002} found such a condition to be met for buoyancy defined relative to a relatively elaborate reference vortex state including even some degree of asymmetry. However, \citet{Zhang2000} found the parcels to be negatively buoyant for buoyancy defined relative to a rest state, the desired upward acceleration being then entirely provided by the pressure gradient term $\nabla p'$.
\end{comment} 

Even if (\ref{buoyant_condition}) holds, it is not sufficient to ensure that $b_e^{(r)}$ be negative. Indeed, because $(\mu-\mu_m)/r^3>0$, (\ref{radial_be}) imposes a further constraint on the magnitude of positive buoyancy anomalies, namely:
\begin{equation}
        \nu_h - \nu_m > \left [ \rho_m \left ( f + \frac{v_m}{r} \right ) v_m \right ]^{-1} \frac{\mu-\mu_m}{r^3} . 
        \label{condition_on_buoyancy} 
\end{equation}
If specific volume anomalies $\nu_h - \nu_m$ are bounded, as is presumably the case in reality, (\ref{condition_on_buoyancy}) appears to impose an upper limit on the maximum angular momentum anomalies $\mu-\mu_m$ and hence on the maximum intensity that the
\textcolor{violet}{unbalanced part of the} vortex can reach. 
\begin{comment} 
This limit is {\em a priori} different from the maximum potential intensity (MPI) predicted by \citet{Emanuel1986} (see \citet{Emanuel2018} for a recent review on this topic and wider TC research), which is reached when the production of available energy by surface enthalpy fluxes balances dissipation by surface friction in the region of maximum winds. Whether such a condition could account for why the intensity of many observed TCs remain significantly below their theoretical maximum intensity \citep{Emanuel2000} is left for future study.
\end{comment} 
\textcolor{violet}{Eqs. (\ref{secondary_circulation}) or (\ref{secondary_circulation_bis}) improve upon Eqs (8) and (9) of \citet{Montgomery2017} by emphasising that the generalised buoyancy force is really a force primarily acting on ${\bf u}_a$ rather than ${\bf u}_s$, as well as by providing more details on the nature of the forces and other effects controlling the secondary circulation.} 

%To the extent that it is relevant, the constraint (\ref{condition_on_buoyancy}) is %interesting because it does not rely on any dissipative mechanism. This is in %contrast with the prevailing WISHE (Wind Induced Surface Heat Exchange) paradigm of %TC intensification (Emanuel, 1986), which assumes that the maximum intensity of a TC %is reached when the energy dissipated by surface friction becomes large enough to %balance the production of APE by the surface enthalpy fluxes. Further work is %required to determine whether (\ref{condition_on_buoyancy}) might be a less %restrictive constraint on TC intensity than that imposed by surface friction, which %would explain why it does not appear to have been discussed before in the TC %literature. Finally, note that a positive buoyancy anomaly $\nu_h - \nu_m >0$ %implies from (\ref{vertical}) a positive upward acceleration $b_e^{(z)}>0$, which %appears to be consistent with the observations of strong updrafts in the  eyewall %region of TCs. 

\textcolor{violet}{To characterize the energetics of the unbalanced part of the vortex intensification,} we find that the simplest and most economical description is one based on separate evolution equations for: the sum of the kinetic energy of the secondary circulation plus the AAE, ${\bf u}_s^2/2 + \Pi_1$; the eddy slantwise APE $\Pi_e$; and the eddy mechanical energy $\Pi_k$. This leads to the following set of equations:

\begin{figure}
    \centering
    \includegraphics[width=13cm]{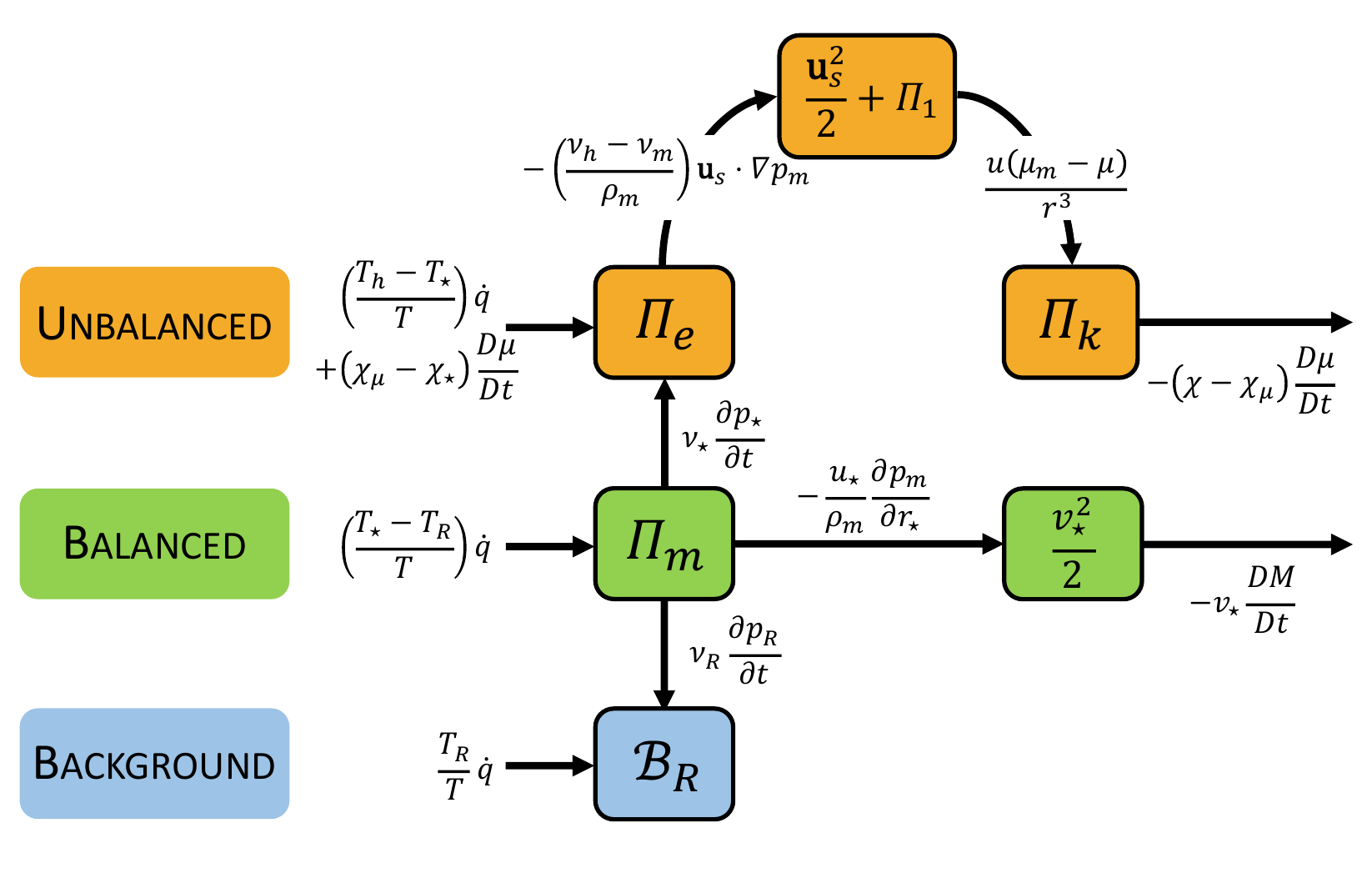}
    \caption{Hypothesised energy pathways associated with the intensification of a cyclonic vortex forced by sources of diabatic heating $\dot{q}$ and angular momentum $D\mu/Dt$.}
    \label{fig:energy_cycle}
\end{figure}

\begin{equation}
    \rho \frac{D}{Dt} \left ( \frac{{\bf u}_s^2}{2} 
    + \Pi_1 \right ) + \nabla \cdot (p'{\bf u}_s) =
     \rho ( {\bf b}_e^T \cdot {\bf u}_s + {\bf b}_e^M \cdot {\bf u}_s ) 
     + \rho G_s , 
\end{equation}
\begin{equation}
    G_s = \left ( \frac{T-T_h}{T} \right ) \dot{q}
    + \nu' \left ( \frac{\partial p_m}{\partial t}
    + {\bf u}_s \cdot\nabla p_m \right ) , 
\end{equation}
\begin{equation}
    \frac{D\Pi_e}{Dt} = - {\bf b}_e^T \cdot {\bf u}_s + 
    \left ( \frac{T_h - T_{\star}}{T} \right ) \dot{q} 
    + (\chi_{\mu}-\chi_{\star}) \frac{D\mu}{Dt} 
    + \nu_h \frac{\partial p_m}{\partial t} - \nu_{\star}
    \frac{\partial p_{\star}}{\partial t} ,
    \label{pie_equation} 
\end{equation}
\begin{equation}
    \frac{D\Pi_k}{Dt} = - {\bf b}_e^M \cdot {\bf u}_s + \left ( \chi - \chi_{\mu} \right ) \frac{D\mu}{Dt} . 
    \label{pik_equation}
\end{equation}
For an intensifying vortex resulting from an increase in $v'$, we established in the previous section that $\nu_h-\nu_m>0$ and $\mu-\mu_m>0$. The implications for the work against the generalised inertial and buoyancy forces ${\bf b}_e^T$ and ${\bf b}_e^M$ by the secondary circulation are:
\begin{equation}
    -{\bf b}_e^M\cdot {\bf u}_s = (\mu-\mu_m) \nabla \chi \cdot {\bf u}_s = 
    - \frac{u(\mu-\mu_m)}{r^3} > 0  , 
\end{equation}
\begin{equation}
    -{\bf b}_e^T \cdot {\bf u}_s = (\nu_h-\nu_m) \nabla p_m \cdot {\bf u}_s = (\nu_h - \nu_m) \left [ u \frac{\partial p_m}{\partial r} - \rho_m g w \right ] < 0 . 
\end{equation}
The sign of such energy conversions suggest that the flow of energy follows the paths
\begin{equation}
      \Pi_e \quad \rightarrow \quad \frac{{\bf u}_s^2}{2} + \Pi_1 
      \quad \rightarrow \quad \Pi_k ,
\end{equation}
as illustrated in Fig. \ref{fig:energy_cycle}. If we neglect the terms related to the time-dependence, the following term needs to be positive
\begin{equation}
      \left ( \frac{T_h - T_{\star}}{T} \right ) \dot{q} 
      + (\chi_\mu - \chi_{\star}) \frac{D\mu}{Dt} > 0 .
      \label{source_term}
\end{equation}
If $D\mu/Dt <0$ acts as a retarding effect, Fig. \ref{fig:lifting_illustration} shows that $(r_{\star}-r_\mu)>0$ and hence that $(\chi_\mu-\chi_{\star})>0$, suggesting that the sink of angular momentum is of the wrong sign. Therefore, for (\ref{source_term}) to act as a source term, the diabatic term must be positive and larger than the term proportional to the angular momentum sink term, viz.,  
\begin{equation}
       \left ( \frac{T_h - T_{\star}}{T} \right ) \dot{q} > 
      \left | (\chi_\mu - \chi_{\star} ) \frac{D\mu}{Dt} \right |  > 0 . 
      \label{thermo_criterion} 
\end{equation}
By definition, $T_h = T(\eta,p_m)$ and $T_{\star} = T(\eta,p_{\star})$, so again from Fig. \ref{fig:lifting_illustration}, $p_m - p_{\star}>0$ and therefore $T_h-T_{\star}>0$. Now, if we regard $p_m = \hat{p}_m(\eta_m,\mu_m)$ as a function of the reference entropy and squared angular momentum, we have
\begin{equation}
      \frac{(T_h - T_{\star})}{T} \approx \frac{1}{T} 
      \frac{\partial T}{\partial p} (p_m - p_{\star})
      \approx - \frac{1}{T} \frac{\partial T}{\partial p} 
      \left \{ \frac{\partial \hat{p}_m}{\partial \eta_m} (\eta-\eta_m)
       + \frac{\partial \hat{p}_m}{\partial \mu_m} (\mu - \mu_m) + \cdots \right \} .
\end{equation}
Since in general pressure varies little with $\mu_m$, it follows that the term is dominated by the entropy anomaly, which needs to be positive as $\partial \hat{p}_m/\partial \eta_m< 0$. For the intensification of $v'$ to proceed, a finite-amplitude entropy anomaly $\eta'$ needs to be produced in order to make the thermodynamic efficiency $(T_h-T_{\star})/T$ large enough to satisfy the threshold relation (\ref{thermo_criterion}).

\section{Discussion and conclusions}

\textcolor{violet}{Adiabatic and inviscid axisymmetric perturbations to a compressible stratified vortex are strongly controlled by the two stabilising factors arising from the material invariance of entropy and angular momentum: static stability, which provides resistance to upward motions, and inertial stability, which provides resistance to radial motions \citep{Eliassen1952}. As a result, such perturbations tend to be oscillatory in nature and made up of internal, inertial/centrifugal and acoustic waves, as is well known, e.g., \citet{Emanuel1994}.}  \textcolor{blue}{For incompressible axisymmetric motions, \citet{Cho1993} hypothesised that this should make it possible to extend Lorenz's available potential energy by augmenting it with the centrifugal potential energy. }\textcolor{violet}{In this paper, we demonstrated their proposition for the first time for the more general case of a compressible fluid. An important outcome is a partitioning of available energy into available acoustic available energy, slantwise available potential energy and centrifugal potential energy that is both different and much simpler than that discussed in \citet{Codoban2006} or \citet{Andrews2006}.}

\textcolor{violet}{Although \citet{Codoban2003,Codoban2006} and \citet{Andrews2006} have previously described the conditions for symmetric stability as determined by three mathematical inequalities, the two-dimensional nature of the problem suggests that only two inequalities are sufficient. We proved this here for the first time for finite-amplitude perturbations and showed that the two inequalities sufficient for determining stability reduce to $N_{slantwise}^2>0$ and $I^2 = r^{-3}\partial \mu/\partial r>0$, where $N_{slantwise}^2$ is the squared buoyancy frequency associated with slantwise displacements along surfaces of constant angular momentum and $I^2$ the inertial stability parameter, the partial derivative $\partial \mu/\partial r$ being taken at constant pressure. These conditions are significantly more restrictive than the conditions $N^2>0$ and $I^2>0$, where $N^2$ is the traditional squared buoyancy frequency associated with upward displacements, as they preclude the possibility of conditional symmetric instability (CSI), whereas the latter do not.} \textcolor{blue}{Although these conditions are not really new, we believe that they are significantly clearer than in previous work. 
}

\textcolor{violet}{A key aspect of adiabatic and inviscid axisymmetric evolution is that the balanced and unbalanced parts of a vortex (which are regarded here as a particular form of mean/eddy partitioning) obey separate energy conservation principles, as first established by \citet{Codoban2003,Codoban2006}, \citet{Andrews2006} and confirmed here. Such a property amounts to some form of non-interaction principle between the balanced and unbalanced parts that places strong constraints on admissible energy conversions. As a result, axisymmetric vortex evolution forbids the conversion of mean APE into eddy energies, which are the signature of baroclinic instability and a prominent feature of the Lorenz energy cycle in the atmosphere. It also forbids the back-scatter of eddy energies into mean energies that would potentially allow for the rectification of the balanced part of the vortex by its unbalanced part. From the viewpoint of energetics at least, the present results definitely confirm that axisymmetric vortex evolution should be regarded as fundamentally different from asymmetric evolution, which has been a longstanding issue in the TC literature at and the centre of the rotating convection paradigm, see \citet{Persing2013,Montgomery2014,Montgomery2017}.
}

\textcolor{violet}{Our framework improves upon that of \citet{Andrews2006} by fully accounting for the effects of sinks and sources of entropy and angular momentum. A key result is that 
in the same way that thermodynamic efficiencies can be defined to predict the fraction of the thermodynamic sources/sinks of energy creating mean and eddy available potential energies, mechanical efficiencies can be similarly defined that predict the fraction of the mechanical sinks/sources of energy creating mean and eddy centrifugal potential energies, as was first shown by \citet{Codoban2003} for zonal Boussinesq flows.
To illustrate the potential usefulness of our framework, we sought to clarify the issue of how a positive source of diabatic heating and sink of angular momentum cooperate to intensify an incipient warm core cyclonic vortex, which is motivated by the problem of TC intensification. As expected, the positive diabatic heating is found to be the ultimate source of energy for both the intensification of the balanced and unbalanced parts of the vortex, but with important differences in the energy conversions involved. In both cases, positive diabatic heating first creates available potential energy, but with different thermodynamic efficiencies. For the balanced part, the APE generated is associated with the system buoyancy and is directly converted into the kinetic energy of the balanced azimuthal circulation, with friction being essential for promoting such a conversion. For the unbalanced part, the APE generated is associated with the generalised buoyancy/inertial forces, and is partly converted into the kinetic energy of the secondary circulation, which is then converted into the kinetic energy of the unbalanced part of the azimuthal circulation, with friction acting as a brake and a constraint on the unbalanced intensification. These results are synthesised in our Fig. \ref{fig:energy_cycle}. Our framework makes it clear that the generalised buoyancy force first discussed by \citet{Smith2005} only pertains to the intensification of the unbalanced part of the vortex, which does not appear to have been clearly acknowledged in the literature. Indeed, the importance of the generalised buoyancy force has been primarily discussed in the context of purely balanced models of TC evolution by \citet{Smith2005}, who showed that the Sawyer-Eliassen equations are partly controlled by the rate of generation of the generalised buoyancy force. \citet{Smith2005}'s result is important because it establishes that purely balanced models of TC evolution are actually capable of capturing aspects of the physics of both balanced and unbalanced intensification; however, it could easily be misinterpreted as suggesting that the generalised buoyancy force is relevant to the intensification of the balanced part of the vortex, which our results establish is not the case.
}

\textcolor{violet}{The partitioning of an axisymmetric vortex into balanced and unbalanced parts plays a central role in our approach, as it does in many studies of TC intensification. In this paper, we took the view that there is really only one physical situation for which such a partitioning is a priori well posed, namely that for which the unbalanced part can be regarded as an inviscid and adiabatic perturbation to a symmetrically and inertially stable balanced part. Such a view motivates regarding the balanced part of a vortex as a state of minimum energy obtainable by means of an inviscid and adiabatic re-arrangement of mass, as underlies the works of \citet{Cullen2015} or \citet{Methven2015}, thus naturally extending \citet{Lorenz1955}'s construction of the background reference state in the theory of available potential energy. Although such a reference state is challenging to construct in practice, it is relatively easy to study mathematically and allows for several definite theoretical statements about the evolution of both the balanced and unbalanced parts of the vortex. For example, it guarantees that the evolution of the purely balanced part of the vortex is diabatic in nature, and fully determined at all times by the knowledge of the diabatic sinks and sources of angular momentum and entropy. Moreover, it leads to a view of energetics in which the balanced and unbalanced parts have distinct signatures. This is in contrast to existing approaches, such as that of \citet{Smith2006}, which are relatively easy to implement in practice as diagnostics, but which are otherwise hard to study theoretically.  
}

\textcolor{violet}{Despite accounting for the effects of moisture only indirectly through the diabatic heating term $\dot{q}$, our dry framework nevertheless exhibits many key features of existing theories of TC intensification, which suggests that it could potentially be useful for understanding how to recast such theories within a single unifying framework. Indeed, existing paradigms of TC intensification tend to be presented as mutually exclusive or rooted in different assumptions that are not necessarily clearly stated even if efforts to improve the situation have been made, e.g., \citet{Montgomery2014}. What our framework suggests, however, is that while a given paradigm of TC intensification may fail on its own, some of the ideas on which it relies may nevertheless be useful or valid for shedding light on particular aspects of the problem. For instance, one of our main results is that friction appears essential for the intensification of the balanced part of the vortex while at the same time acting as a sink of energy for both the balanced and unbalanced parts; this therefore vindicates \citet{Charney1964}'s idea that friction should be regarded as having the dual role of both promoting and acting against TC intensification, which is central to the otherwise refuted CISK (Conditional Instability of the Second Kind) paradigm. On the other hand, our framework appears to support the heat engine view of TCs, although the thermodynamic efficiencies controlling the rate of APE production appear to be somewhat different from those entering potential intensity theories (see \citet{Montgomery2017} for a review of these).}

\textcolor{violet}{Moreover, because such thermodynamic efficiencies appear to depend on the state of the system and therefore potentially functions of time, it makes it possible for TC intensification, at least in principle, to occur due to an increasing thermodynamic efficiency rather than via the WISHE (Wind Induced Surface Heat Exchange) feedback. This is similar to the argument of \citet{Schubert1982} that the rapid intensification of TCs could be caused by increases in efficiency. The applicability of this theory was questioned by \citet{Smith2016}, who highlighted that the efficiency as defined by \citet{Schubert1982} is unable to account for the effects of boundary layer dynamics or thermodynamics, which are known to be crucial for intensification. It is possible that using an alternative definition of efficiency, such as the ones developed here, could remedy the shortcomings of the study by \citet{Schubert1982} whilst still demonstrating the importance of the temporal development of efficiency for the TC. This is consistent with the idea that the WISHE paradigm is unlikely to be the full story for explaining TC intensification in axisymmetric models \citep{Montgomery2009DoWISHE, Montgomery2015, Zhang2016OnIntensification}. }

\textcolor{violet}{Since our framework is capable of addressing both the physics of the balanced and unbalanced parts, it is also potentially helpful for formalising the problem of how to account for so-called superintensities, e.g., \citet{Rousseau-Rizzi2019}. Finally, it raises the question to what extent rectification of the balanced part by the unbalanced part due to the back-scatter of eddy energies into mean energies and/or conversions of mean energies into eddy energies, which are impossible in axisymmetric vortices, account for the observed differences between non-symmetric and axisymmetric TC evolution that form a central part in the rotating convection paradigm, a topic for future work. }

\section*{Acknowledgements}
BLH acknowledges support from NERC as part of the SCENARIO Doctoral Training Partnership (NE/L002566/1). The code used to produce the illustrations of the available energetics of an axisymmetric vortex in a dry atmosphere is available at https://github.com/bethanharris/vortex-available-energy. \textcolor{violet}{The authors thank T.G. Shepherd, R.K. Smith and M. T. Montgomery for comments and references to the literature that greatly helped clarify several important issues.} Declaration of interest. The authors report no conflicts of interest. 

\appendix

\section{Analytical expression for vortex motions}
\label{analytical_solution} 

Many of the illustrations of this paper are based on a dry idealised tropical cyclone axisymmetric vortex taken from \citet{Smith2005}, the description of which is reproduced here. 
This idealised TC is defined by its pressure perturbation 
\begin{equation}
    p(s,z) = (p_c - p_{\infty}(0)) \left [ 1 - \exp{\left ( \frac{-x}{s} \right )} \right ]  \exp{\left ( \frac{-z}{z^{\star}} \right )} 
    \cos{\left ( \frac{\pi}{2} \frac{z}{z_0} \right )} ,
\end{equation}
where $p_c$ is the central pressure at the surface, $p_{\infty}(0)$ is the surface pressure at large radial distance, $s=r/r_m$ and $x$, $r_m$, $z_0$ and $z^{\star}$ are constants. We choose $p_{\infty}(0) - p_c$ and $x$ so that the maximum tangential wind speed is about $40\,{\rm m}\cdot {\rm s}^{-1}$ at a radius of about 40 km and declines to zero at an altitude $z_0 = 16\,{\rm km}$: specifically $p_{\infty}(0)-p_c = 50\,{\rm mb}$, $z^{\star} = 8 \,{\rm km}$ and $x=1.048$. The exponential decay with height approximately matches the decrease in the environmental density with height and is necessary to produce a reasonably realistic tangential wind distribution that decreases in strength with height.

\bibliography{jfm-references}
\bibliographystyle{jfm}

\end{document}